\pdfoutput=1
\PassOptionsToPackage{table}{xcolor}

\documentclass[11pt]{article}

\usepackage[final]{acl}

\usepackage{times}
\usepackage{latexsym}
\usepackage{amsmath}
\usepackage{amsfonts}
\usepackage{graphicx}
\usepackage{xcolor}
\usepackage{pgfplots}
\pgfplotsset{compat=1.18}

\usepackage[T1]{fontenc}

\usepackage[utf8]{inputenc}

\usepackage{microtype}
\usepackage{booktabs}
\definecolor{mmdrow}{HTML}{EAF2FF}
\definecolor{gaincol}{HTML}{EEF7F1}
\usepackage{inconsolata}

\title{An Embarrassingly Simple Detector for Model Extraction Attacks \\in Large Language Model API Traffic}

\author{
Shuze Liu \\
Florida State University \\
\texttt{sl26u\symbol{64}fsu.edu}
\And
Qianwen Guo \\
Florida State University \\
\texttt{qguo\symbol{64}fsu.edu}
\And
Yushun Dong \\
Florida State University \\
\texttt{yd24f\symbol{64}fsu.edu}
}

\begin{document}
\maketitle
\begin{abstract}
Large language models (LLMs) are increasingly deployed through hosted APIs, making model extraction a practical threat to model ownership and service security.
Individual extraction queries often resemble benign requests, while existing evaluations often focus on single-query anomaly scoring or pure benign-versus-attacker user settings.
We formulate model extraction monitoring as benign-calibrated traffic-window distribution testing: embed incoming queries into a semantic space and test whether their aggregate distribution deviates from historical benign traffic.
We instantiate this formulation with maximum mean discrepancy (MMD), using only benign-vs-benign comparisons to set the decision threshold.
We evaluate on fourteen attacker-normal query pairs from four extraction scenarios and compare with adapted PRADA, SEAT, CAP, DATE, marginal Mahalanobis, and pseudo-class energy baselines.
Across three random seeds, MMD achieves 0.3\% benign FPR, 100.0\% pure-attacker TPR, 90.5\% average TPR over attacker fractions, and 95.1\% balanced accuracy.
These results show that benign-calibrated distribution testing is a strong empirical baseline for model extraction detection in both user-level and mixed multi-user LLM API traffic.
Code is released at: \url{https://github.com/LabRAI/mmd-llm-mea-detection}.
\end{abstract}

\section{Introduction}

Large language models (LLMs) are increasingly deployed through hosted APIs, allowing users to access powerful models without direct access to their parameters \citep{brown2020language,bommasani2021opportunities}.
This deployment model protects model weights, but it does not eliminate model extraction risks.
Classical model stealing attacks show that prediction APIs can be repeatedly queried to train substitute models \citep{tramer2016stealing,papernot2017practical,orekondy2019knockoff,jagielski2020high,chandrasekaran2020exploring}, and recent surveys identify model extraction as a major threat for LLM services \citep{zhao2025surveyllmextraction,zhao2025systematicsurvey}.
In NLP, attackers extract BERT-based APIs using synthetic text queries \citep{krishna2020thieves}, and recent work has extended extraction to LLM capabilities, domain knowledge, and even parts of production language models \citep{birch2023modelleeching,dai2023meaeq,li2026queryefficient,carlini2024stealing}.
Because these attacks require many target-service queries, query-stream monitoring can catch extraction early, before a full substitute model is completed.

However, deployment-oriented model extraction monitoring is not fully captured by existing evaluation settings.
Embedding-space OOD methods such as Mahalanobis and energy-based scoring measure deviation at the level of individual inputs \citep{lee2018simple,liu2020energy}, and text anomaly detectors such as DATE similarly score each query \citep{manolache2021date}.
Existing model extraction detectors are commonly evaluated at the user or account level, where a benign user issues only legitimate queries and an attacker user executes a complete extraction workflow \citep{juuti2019prada,zhang2021seat,kulkarni2026stealing}.
Other defenses rely on chronological account-level streams, task-specific encoders, self-supervised anomaly models, or active output perturbation tied to a particular extraction setting \citep{juuti2019prada,zhang2021seat,manolache2021date,kulkarni2026stealing}.
These settings capture important aspects of model extraction defense, but they do not directly address passive, query-only monitoring over benign-calibrated traffic windows that may mix benign and attacker queries.

This gap creates three practical challenges.
\textit{Challenge 1: Query-level ambiguity.}
Individual attack queries often appear benign because extraction methods draw from natural text sources, task inputs, or domain-specific templates, such as Wikipedia-like text, SQuAD prompts, and medical knowledge questions \citep{krishna2020thieves,birch2023modelleeching,dai2023meaeq,li2026queryefficient}.
The relevant signal is therefore a weak distributional shift induced by repeated extraction queries rather than a clearly anomalous single query.
\textit{Challenge 2: Mixed-traffic dilution.}
In aggregate API monitoring, incoming traffic may combine multiple users, so attacker queries can appear as a small fraction of a larger, potentially cross-user window.
Evaluation only on pure benign users versus pure attacker users does not fully address this mixed-traffic setting.
\textit{Challenge 3: Benign-calibrated low-FPR operation.}
Defenders typically have historical benign traffic but lack the attacker's query generator and labeled attack examples, while large alarm volumes create analyst burden and reduce detector usability \citep{julisch2003clustering,layman2023controlled}.
Thus, a deployment-oriented protocol should calibrate from benign queries alone and evaluate detection under both pure and mixed attack traffic, while reporting benign false alarms explicitly.

To tackle these challenges, we formulate model extraction monitoring as benign-calibrated traffic-window distribution testing.
The central question is whether repeated extraction queries induce a measurable shift in the semantic distribution of an incoming traffic window, even when individual queries look plausible and are diluted by benign requests.
We instantiate this formulation by encoding queries with a sentence embedding model and comparing each incoming traffic window against benign reference traffic with maximum mean discrepancy (MMD), a kernel two-sample statistic \citep{gretton2012kernel}.
Rather than assigning each query an independent deviation score and aggregating afterward, this two-sample view directly compares incoming and benign-reference distributions.
The threshold is calibrated using only benign-vs-benign comparisons, making the detector independent of labeled attack data.
We evaluate this formulation on fourteen attacker-normal query pairs from four extraction scenarios, covering pure attacker traffic and mixed multi-user traffic, and compare it with PRADA, SEAT, CAP, DATE, marginal Mahalanobis distance, and pseudo-class energy.

The main contribution of this paper is summarized as follows.
(1) \textbf{Deployment-oriented problem formulation.}
We formulate model extraction query detection as benign-calibrated traffic-window distribution testing for hosted LLM APIs, covering both pure attacker-user traffic and mixed, potentially cross-user traffic where extraction queries may be diluted by benign requests.
This setting is not directly addressed by prior single-query anomaly detection or pure benign-versus-attacker user evaluations, which assume a different decision unit.
(2) \textbf{Window-level detector and unified deployment protocol.}
We instantiate this formulation with an attack-label-free MMD two-sample detector and adapt six extraction, anomaly, and OOD baselines to the same query-only embedding and benign-calibration protocol.
(3) \textbf{Low-FPR empirical finding.}
Across fourteen attacker-normal pairs, MMD achieves 0.3\% benign FPR, 100.0\% pure-attacker TPR, 90.5\% Avg. TPR, and the highest balanced accuracy among evaluated methods, showing that a simple two-sample traffic-window detector provides a strong low-FPR baseline for this practical monitoring problem.

\section{Preliminaries and Problem Definition}

\paragraph{Preliminaries.}
Let $\mathcal{Q}$ denote the space of natural-language queries submitted to a hosted language-model service.
The target service is represented as $F$, which receives a query $q \in \mathcal{Q}$ and returns a model response.
In a model extraction attack, an adversary repeatedly queries $F$ to collect input-output observations for approximating the target model or its capabilities \citep{tramer2016stealing,orekondy2019knockoff}.
The defender observes the incoming queries but does not assume access to the attacker's generation process, private model, or attack labels during deployment.
We denote the benign query distribution as $P_b$ and the attacker query distribution as $P_a$.
The defender has access to a benign historical query set $B=\{q_i^b\}_{i=1}^{N_b}$ sampled from $P_b$.
At test time, the defender receives an incoming traffic window, represented as a query batch $T=\{q_i^t\}_{i=1}^{N_t}$.
The window can correspond to queries from one user account, a group of accounts, or an aggregate API traffic stream over a short monitoring interval.
A benign batch is sampled from $P_b$, while an attack batch is modeled as a contaminated distribution $(1-\rho)P_b+\rho P_a$, where $\rho \in [0,1]$ is the attacker fraction.
The pure attack case corresponds to $\rho=1$, and mixed attack cases correspond to $0<\rho<1$.
We therefore detect distributional shifts over traffic windows rather than labeling individual queries, which avoids requiring query-level attack annotations at deployment time.

To compare query batches, we map each query into a semantic space using a fixed encoder $\phi:\mathcal{Q}\rightarrow\mathbb{R}^d$.
The benign reference set and incoming batch are represented as $Z_B=\{\phi(q):q\in B\}$ and $Z_T=\{\phi(q):q\in T\}$.
The detector computes a statistic $s(Z_T,Z_B)$ that measures deviation from the benign reference distribution.
As in PRADA, repeated-query distributional deviation is our detection signal \citep{juuti2019prada}.

\paragraph{Problem 1. Traffic-window model extraction query detection.}
Given a benign reference set $B$ and an incoming query batch $T$, the goal is to learn or calibrate a detector $h$ that outputs
\begin{equation}
    h(T;B) \in \{0,1\},
\end{equation}
where $h(T;B)=1$ indicates that $T$ is suspicious.
The detector should maintain a low false positive rate when $T\sim P_b$ while achieving a high detection rate when $T$ contains attacker queries.
Formally, for a target false positive level $\alpha$, we aim to maximize detection power under attack contamination,
\begin{equation}
    \max_h \ \Pr[h(T;B)=1 \mid T\sim (1-\rho)P_b+\rho P_a],
\end{equation}
subject to
\begin{equation}
    \Pr[h(T;B)=1 \mid T\sim P_b] \leq \alpha.
\end{equation}
We calibrate the detector using benign-vs-benign comparisons from $B$ and evaluate it on pure and mixed attack batches in testing.

\section{Methodology}

In this section, we first present an overview of our MMD-based detection framework, followed by the detailed elaboration on its two main components: embedding-space query distribution modeling and benign-calibrated MMD detection.
Finally, we summarize offline calibration and online detection.

\subsection{Overview}

\begin{figure*}[t]
    \centering
    \includegraphics[width=\textwidth,trim=0in 2.3in 0in 0.1in,clip]{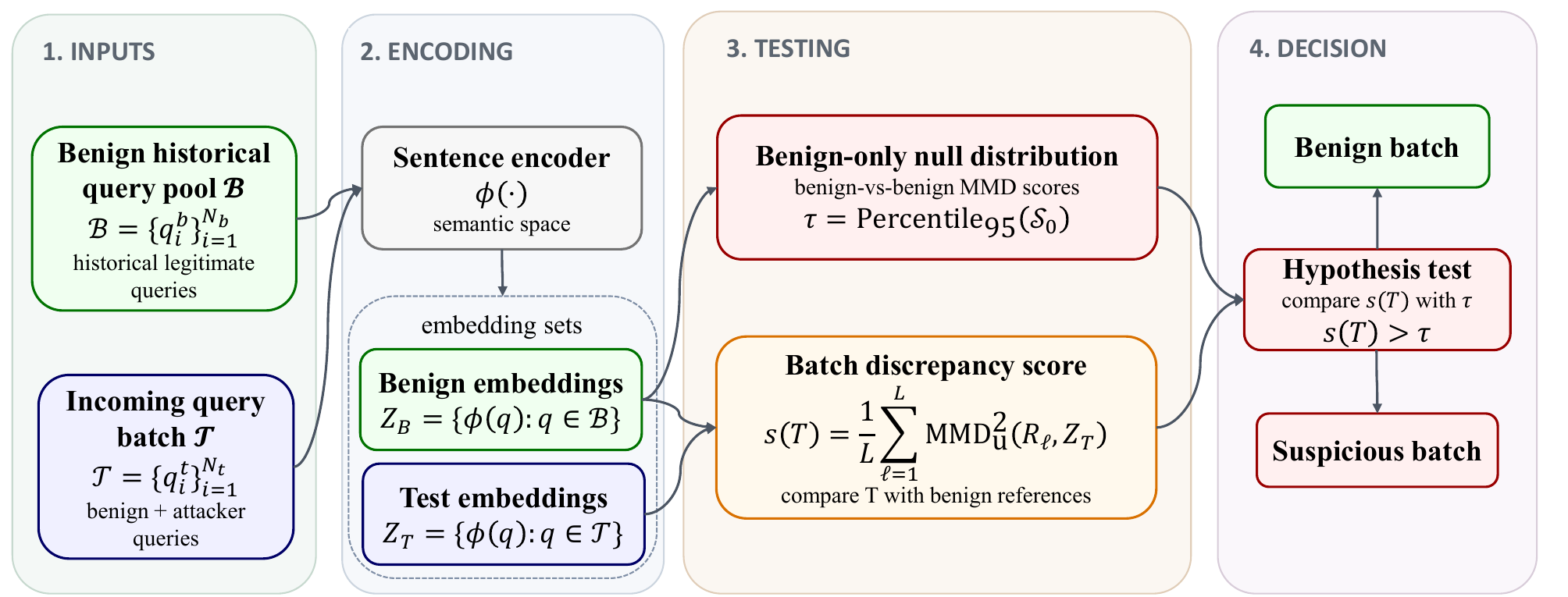}
    \caption{Overview of the proposed benign-calibrated query-traffic detection framework.}
    \label{fig:method_overview}
\end{figure*}

We introduce the workflow of the proposed framework in Fig.~\ref{fig:method_overview}.
The detector monitors a traffic window rather than isolated queries.
It embeds historical benign queries and incoming queries into the same semantic space, compares their empirical distributions, and calibrates the alarm threshold using only benign traffic.

The detection pipeline contains three stages.
First, the defender builds a benign reference pool from historical normal queries and embeds the queries with a fixed sentence encoder.
Second, the defender constructs a benign null distribution by repeatedly comparing two benign batches with MMD and sets the threshold as a high percentile of this distribution.
Third, when an incoming traffic window arrives, the detector computes its MMD discrepancy against benign reference batches and flags the window if the score exceeds the threshold.
This design needs no labeled attack queries and applies to pure attacker and mixed traffic under the same online decision rule.

\subsection{Query Distribution Modeling}

We first transform raw text queries into a continuous semantic representation.
Directly comparing raw queries is brittle because extraction queries use natural text, task inputs, or templates whose surface forms resemble benign user requests, despite serving an extraction objective.
Let $\phi:\mathcal{Q}\rightarrow\mathbb{R}^d$ denote a fixed sentence encoder.
Given a benign query set $B=\{q_i^b\}_{i=1}^{N_b}$ and an incoming query batch $T=\{q_i^t\}_{i=1}^{N_t}$, we obtain the embedded sets
\begin{equation}
    Z_B=\{\phi(q):q\in B\}, \qquad
    Z_T=\{\phi(q):q\in T\}.
\end{equation}
The detector treats $Z_B$ as samples from the benign query distribution and $Z_T$ as samples from the incoming batch distribution.
When the incoming batch contains extraction queries, its aggregate embedding distribution exhibits measurable shifts away from the benign reference distribution even when individual queries appear plausible.

To reduce variance caused by a particular reference sample, the detector does not compare $Z_T$ against a single fixed benign batch.
Instead, it samples multiple benign reference batches $R_1,\ldots,R_L$ from $Z_B$, each with the same batch size as the incoming batch.
The final score of $T$ is obtained by averaging its discrepancy against these reference batches.
This reference averaging reduces sensitivity to random benign sampling noise and makes the score less dependent on any single benign batch.

\subsection{MMD Detection}

We instantiate the batch discrepancy with maximum mean discrepancy (MMD), a kernel two-sample statistic for comparing two empirical distributions \citep{gretton2012kernel}.
Given two embedding batches $X=\{x_i\}_{i=1}^{m}$ and $Y=\{y_j\}_{j=1}^{n}$, and a positive definite kernel $k(\cdot,\cdot)$, the unbiased squared MMD estimator is
\begin{equation}
\begin{aligned}
\mathrm{MMD}_u^2(X,Y)
=&\frac{1}{m(m-1)}\sum_{i\neq j}^{m} k(x_i,x_j) \\
&+\frac{1}{n(n-1)}\sum_{i\neq j}^{n} k(y_i,y_j) \\
&-\frac{2}{mn}\sum_{i=1}^{m}\sum_{j=1}^{n} k(x_i,y_j).
\end{aligned}
\end{equation}
We use an RBF kernel over query embeddings.
The base bandwidth is selected with the median heuristic on benign embeddings, and we average multiple RBF kernels with bandwidths scaled by $\{0.5,1,2,4\}$.
This multi-kernel design reduces sensitivity to a single bandwidth choice.

For an incoming batch $T$, the detector samples $L$ benign reference batches $R_1,\ldots,R_L$ from $Z_B$ and computes the averaged discrepancy score
\begin{equation}
    s(T)=\frac{1}{L}\sum_{\ell=1}^{L}\mathrm{MMD}_u^2(R_\ell,Z_T).
\end{equation}
A larger score indicates that the incoming batch is farther from the benign reference distribution in embedding space.
To calibrate a threshold without attack data, we build a benign null distribution by sampling pairs of benign batches from $Z_B$:
\begin{equation}
    \mathcal{S}_0=\{\mathrm{MMD}_u^2(R_a^{(k)},R_b^{(k)})\}_{k=1}^{K}.
\end{equation}
The threshold $\tau$ is set to the $\gamma$-th percentile of $\mathcal{S}_0$, where $\gamma=95$ in our experiments as a conservative low-FPR default.
The final decision rule is
\begin{equation}
    h(T;B)=\mathbf{1}[s(T)>\tau].
\end{equation}
We also compute an empirical p-value by comparing $s(T)$ with the benign null scores.

\subsection{Detection Strategy}

The detector has an offline calibration stage and an online detection stage.
During offline calibration, historical benign queries are split into a benign reference pool and a benign test pool.
The reference pool is used to estimate kernel bandwidths, sample reference batches, and construct the benign null distribution.
The remaining benign test pool is not used to choose the threshold; it is reserved for estimating false positive rate.

During online detection, the defender collects a batch of incoming queries, embeds them with the same encoder, computes the averaged MMD score against sampled benign reference batches, and compares the score with the benign-calibrated threshold.
The output is a batch-level suspicious flag rather than query-level labels, which matches how operators typically triage traffic-level alerts.
This matches the operational setting where model extraction attacks require repeated queries, while each individual query remains ambiguous.

For evaluation, we apply the same detection strategy to three types of batches.
Benign-only batches measure the false positive rate and correspond to normal-user traffic.
Pure attacker batches measure detection performance when the full batch is generated by an extraction attack.
Mixed batches combine benign and attacker queries to test whether aggregate shifts expose weak contamination before attacker traffic dominates the window.

\section{Experimental Evaluations}

In this section, we first introduce the experiment setup.
Then, we discuss the evaluation results of the proposed detector under pure-attacker, mixed-traffic, adaptive, and sensitivity settings.
Specifically, we aim to answer the following research questions:
\textbf{RQ1:} How effective is the proposed detector across diverse model extraction query sources?
\textbf{RQ2:} How does our detector compare with adapted baselines under a unified traffic-window protocol?
\textbf{RQ3:} Does the detector remain effective when attacker queries are mixed with benign traffic at low attacker fractions?
\textbf{RQ4:} How does adaptive evasion affect detection and extraction quality?
\textbf{RQ5:} How sensitive is benign-only calibration to reference-pool contamination?
\textbf{RQ6:} How well do original baseline methods transfer to our model extraction detection setting?
\textbf{RQ7:} How sensitive are results to encoder choice and key practical operating choices?

\begin{table*}[!t]
\centering
\footnotesize
\setlength{\tabcolsep}{2.4pt}
\renewcommand{\arraystretch}{1.12}
\begin{tabular}{lcccccccc}
\toprule
Method & Benign FPR & 5\% TPR & 10\% TPR & 25\% TPR & 50\% TPR & 100\% TPR & Avg. TPR & Balanced Acc. \\
\midrule
\rowcolor{mmdrow}
MMD & \textbf{0.3 $\pm$ 0.3} & 59.0 $\pm$ 1.7 & \textbf{93.7 $\pm$ 2.0} & \textbf{100.0 $\pm$ 0.0} & \textbf{100.0 $\pm$ 0.0} & \textbf{100.0 $\pm$ 0.0} & \textbf{90.5 $\pm$ 0.2} & \textbf{95.1 $\pm$ 0.2} \\
Mahalanobis & 14.2 $\pm$ 9.6 & \textbf{70.5 $\pm$ 6.7} & 83.2 $\pm$ 3.2 & 91.9 $\pm$ 2.7 & 99.0 $\pm$ 0.8 & \textbf{100.0 $\pm$ 0.0} & 88.9 $\pm$ 2.4 & 87.4 $\pm$ 5.6 \\
DATE & 12.5 $\pm$ 7.4 & 39.7 $\pm$ 10.7 & 56.7 $\pm$ 11.8 & 86.1 $\pm$ 5.9 & 95.0 $\pm$ 1.1 & 99.3 $\pm$ 1.2 & 75.4 $\pm$ 4.9 & 81.4 $\pm$ 5.4 \\
SEAT & 13.6 $\pm$ 3.9 & 31.7 $\pm$ 9.5 & 64.5 $\pm$ 5.1 & 93.2 $\pm$ 0.4 & 97.9 $\pm$ 0.8 & \textbf{100.0 $\pm$ 0.0} & 77.5 $\pm$ 2.8 & 82.0 $\pm$ 3.0 \\
Pseudo-energy & 14.3 $\pm$ 8.3 & 54.0 $\pm$ 4.7 & 66.7 $\pm$ 6.7 & 74.5 $\pm$ 4.3 & 78.8 $\pm$ 0.8 & 78.6 $\pm$ 0.0 & 70.5 $\pm$ 3.3 & 78.1 $\pm$ 2.5 \\
CAP & 16.8 $\pm$ 6.3 & 26.4 $\pm$ 3.5 & 34.8 $\pm$ 6.7 & 61.9 $\pm$ 6.4 & 86.0 $\pm$ 3.0 & 95.4 $\pm$ 4.0 & 60.9 $\pm$ 3.3 & 72.0 $\pm$ 1.9 \\
PRADA & 7.3 $\pm$ 3.4 & 18.9 $\pm$ 3.6 & 25.4 $\pm$ 6.6 & 58.2 $\pm$ 2.7 & 66.3 $\pm$ 0.4 & 77.8 $\pm$ 1.2 & 49.3 $\pm$ 2.4 & 71.0 $\pm$ 1.1 \\
\bottomrule
\end{tabular}
\caption{Overall detection results under the unified traffic-window protocol.
Values are averaged over fourteen attacker-normal pairs and three random seeds.
Best results are shown in bold.
Lower is better for benign FPR; higher is better otherwise.
Avg. TPR averages the five attacker fractions.
Balanced Acc. is computed as $(\mathrm{Avg.\ TPR} + 100 - \mathrm{FPR})/2$.}
\label{tab:main_results}
\end{table*}

\subsection{Experiment Settings}

We describe the experiment settings; Appendix~\ref{app:implementation_details} provides additional details.

\paragraph{Datasets.}
We evaluate on fourteen attacker-normal query pairs from four extraction families: \textsc{Query-Efficient-Med} \citep{li2026queryefficient}, \textsc{Model-Leeching} \citep{birch2023modelleeching}, \textsc{MeaeQ} \citep{dai2023meaeq}, and BERT-based API extraction \citep{krishna2020thieves}.
Attacker queries are generated from medical domain exploration, SQuAD-style prompts, or WikiText-103-derived text \citep{merity2016pointer}, while normal queries come from WildChat, SQuAD, GLUE, BoolQ, AG News, Hate Speech, SST-2, and IMDB \citep{zhao2024wildchat,rajpurkar2016squad,wang2018glue,clark2019boolq,zhang2015character,davidson2017automated,socher2013recursive,maas2011learning}.
For fairness, normal queries come from the victim-side task data whenever available; \textsc{Query-Efficient-Med} uses medicine-related WildChat queries because its victim model is GPT.
Appendix~\ref{app:implementation_details} provides the detailed source summary in Table~\ref{tab:dataset_sources}.

\begin{figure*}[!t]
\centering
\definecolor{barMMD}{HTML}{5F9E6E}
\definecolor{barMaha}{HTML}{8AA6C8}
\definecolor{barDATE}{HTML}{E6A35C}
\definecolor{barSEAT}{HTML}{9B83C9}
\definecolor{barEnergy}{HTML}{5CA7A3}
\definecolor{barCAP}{HTML}{D98B8B}
\definecolor{barPRADA}{HTML}{8F969E}
\begin{tikzpicture}
\begin{axis}[
    ybar,
    width=0.96\textwidth,
    height=0.36\textwidth,
    bar width=4pt,
    ymin=0,
    ymax=110,
    ylabel={Rate (\%)},
    symbolic x coords={Specificity,5\% TPR,10\% TPR,25\% TPR,50\% TPR},
    xtick=data,
    ytick={0,25,50,75,100},
    xtick pos=bottom,
    ytick pos=left,
    axis background/.style={fill=gray!3},
    axis line style={gray!70},
    tick style={gray!70},
    grid=major,
    major grid style={line width=.35pt, draw=gray!35},
    tick label style={font=\small},
    label style={font=\small},
    legend columns=7,
    legend style={font=\scriptsize, at={(0.5,-0.25)}, anchor=north, draw=none, /tikz/every even column/.append style={column sep=0.18cm}},
]
\addplot+[fill=barMMD, draw=barMMD] coordinates {(Specificity,99.7) (5\% TPR,59.0) (10\% TPR,93.7) (25\% TPR,100.0) (50\% TPR,100.0)};
\addlegendentry{MMD}
\addplot+[fill=barMaha, draw=barMaha] coordinates {(Specificity,85.8) (5\% TPR,70.5) (10\% TPR,83.2) (25\% TPR,91.9) (50\% TPR,99.0)};
\addlegendentry{Mahalanobis}
\addplot+[fill=barDATE, draw=barDATE] coordinates {(Specificity,87.5) (5\% TPR,39.7) (10\% TPR,56.7) (25\% TPR,86.1) (50\% TPR,95.0)};
\addlegendentry{DATE}
\addplot+[fill=barSEAT, draw=barSEAT] coordinates {(Specificity,86.4) (5\% TPR,31.7) (10\% TPR,64.5) (25\% TPR,93.2) (50\% TPR,97.9)};
\addlegendentry{SEAT}
\addplot+[fill=barEnergy, draw=barEnergy] coordinates {(Specificity,85.7) (5\% TPR,54.0) (10\% TPR,66.7) (25\% TPR,74.5) (50\% TPR,78.8)};
\addlegendentry{Pseudo-energy}
\addplot+[fill=barCAP, draw=barCAP] coordinates {(Specificity,83.2) (5\% TPR,26.4) (10\% TPR,34.8) (25\% TPR,61.9) (50\% TPR,86.0)};
\addlegendentry{CAP}
\addplot+[fill=barPRADA, draw=barPRADA] coordinates {(Specificity,92.7) (5\% TPR,18.9) (10\% TPR,25.4) (25\% TPR,58.2) (50\% TPR,66.3)};
\addlegendentry{PRADA}
\end{axis}
\end{tikzpicture}
\caption{Mixed-traffic detection with benign specificity and attacker TPR.
Benign specificity is $100-\mathrm{FPR}$, so higher values are better.
Other groups report TPR at each mixed attacker fraction.
The 100\% attacker setting is omitted because it is pure attacker-user traffic rather than mixed traffic.}
\label{fig:mixed_traffic_tradeoff}
\end{figure*}
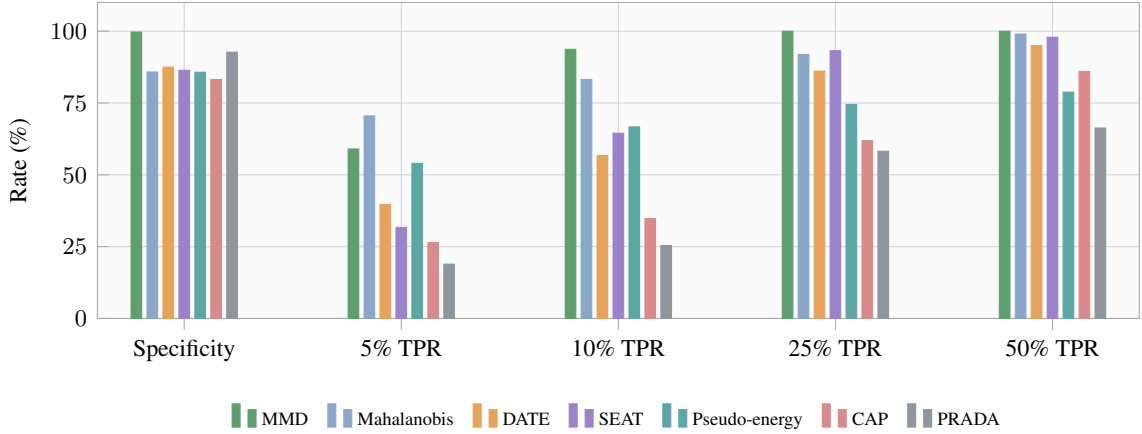
\paragraph{Query encoder.}
We embed all queries with \texttt{BAAI/bge-small-en-v1.5}, a compact BGE/FlagEmbedding sentence embedding model \citep{xiao2023cpack}, and normalize embeddings before scoring in all reported experiments.

\paragraph{Baselines.}
We compare MMD with six adapted baselines: PRADA \citep{juuti2019prada}, SEAT \citep{zhang2021seat}, CAP \citep{kulkarni2026stealing}, DATE \citep{manolache2021date}, marginal Mahalanobis distance \citep{podolskiy2021mahalanobis}, and pseudo-class energy \citep{liu2020energy}.
Because these methods target different settings, we use a unified query-embedding and benign-calibration protocol while preserving each baseline score whenever its original score definition is available.

\paragraph{Evaluation metrics.}
We report true positive rate (TPR), false positive rate (FPR), average TPR over attacker fractions, and balanced accuracy, computed as $(\mathrm{Avg.\ TPR}+100-\mathrm{FPR})/2$.
Benign-only batches represent normal-user traffic, pure attacker batches represent attacker-user traffic, and mixed batches use attacker fractions $\rho\in\{0.05,0.10,0.25,0.50\}$ to model diluted or distributed extraction traffic.

\paragraph{Implementation details.}
For each pair, we split normal queries into an 80\% benign reference pool and a 20\% benign evaluation pool.
Unless otherwise stated, the traffic-window size is 1,500, thresholds are calibrated from 1,000 benign-only calibration samples with the 95th percentile rule, and each setting uses 50 benign-only, 50 pure attacker, and 50 mixed batches.
Main results are averaged over seeds 1, 20, and 42.
For MMD, we use a multi-kernel RBF statistic with 20 benign reference repeats.
Sensitivity experiments vary encoder choice, batch size, threshold, and decision direction.

\subsection{Overall Detection Effectiveness}

To answer RQ1, we evaluate MMD across fourteen attacker-normal pairs under benign-only, pure-attacker, and mixed traffic windows.
Table~\ref{tab:main_results} reports averaged results; Appendix~\ref{app:per_dataset_results} reports detailed per-dataset breakdowns.

From Table~\ref{tab:main_results}, we make the following observations.
(1) MMD separates normal-user and attacker-user traffic almost perfectly, with 0.3\% benign FPR and 100.0\% pure-attacker TPR.
This indicates that benign-only calibration does not create many false alarms while still detecting complete extraction workflows.
(2) It remains effective under mixed traffic, reaching 100.0\% TPR at 25\% and 50\%, 93.7\% at 10\%, and 59.0\% at the hardest 5\% attacker fraction.
(3) Difficulty varies by attack family: template-based or domain-specific settings such as \textsc{Model-Leeching} and \textsc{Query-Efficient-Med} are easier, while WikiText-derived BERT-API settings are harder under low attacker fractions because their attacker queries are semantically close to benign task inputs.
Overall, MMD works for both conventional user-level detection and mixed traffic where attackers insert benign cover queries, slow extraction, or distribute queries across multiple accounts over time in practical deployments.

\subsection{Unified Protocol Comparison}

To answer RQ2, we compare MMD with six adapted baselines under the same traffic-window protocol, using the same semantic representation, benign-only calibration source, traffic-window construction, and evaluation windows.
Because false alarms can overwhelm analysts \citep{julisch2003clustering,layman2023controlled}, benign FPR is central.
From Table~\ref{tab:main_results}, we make the following observations.
(1) MMD gives the best overall trade-off: 0.3\% benign FPR, 100.0\% pure-attacker TPR, 95.1\% Balanced Acc., and Youden's $J$ of 90.2 versus 74.7 for Mahalanobis.
(2) Mahalanobis is highly sensitive to weak low-fraction attacker traffic, achieving the strongest 5\% TPR, 70.5\%, compared with 59.0\% for MMD.
However, this sensitivity also marks many benign windows as anomalous, producing 14.2\% benign FPR.
(3) DATE, SEAT, and pseudo-energy detect many less-diluted attacker windows, with 10\% TPRs of 56.7\%, 64.5\%, and 66.7\%, respectively, but their 5\% attacker TPRs remain below Mahalanobis and their benign FPRs are above MMD.
(4) CAP and PRADA are weaker after adaptation: CAP has the highest FPR, 16.8\%, while PRADA has the weakest Avg. TPR, 49.3\%.
These comparisons highlight why low-FPR operation is central in this deployment setting.
A detector that is more sensitive at very low attacker fractions can still be less useful if it produces many benign alerts, especially when monitoring high-volume API traffic.
Overall, no adapted baseline simultaneously matches MMD's near-zero benign FPR, high attacker TPR, and best Balanced Acc. under the same unified deployment protocol.

\subsection{Mixed and Evasive Traffic}

To answer RQ3, we construct mixed traffic windows where attacker queries occupy $\rho\in\{0.05,0.10,0.25,0.50\}$ of the window and the rest are benign.
This setting corresponds to practical evasive behaviors such as inserting benign cover queries, slowing down extraction over time, or distributing extraction queries across multiple accounts before aggregation.
Figure~\ref{fig:mixed_traffic_tradeoff} summarizes benign specificity, computed as $100-\mathrm{FPR}$, and mixed-traffic TPR under different attacker fractions in this deployment setting.

\begin{figure*}[!t]
\centering
\definecolor{figfpr}{HTML}{8FB7E3}
\definecolor{figfive}{HTML}{E58F7E}
\definecolor{figten}{HTML}{D9A441}
\definecolor{figtwentyfive}{HTML}{78AE6E}
\begin{minipage}{0.49\textwidth}
\centering
\begin{tikzpicture}
\begin{axis}[
    name=batchtpr,
    width=\linewidth,
    height=0.54\linewidth,
    title={(a) Batch size},
    ylabel={TPR (\%)},
    xmin=80,
    xmax=1520,
    ymin=0,
    ymax=105,
    xtick={100,200,500,1000,1500},
    ytick={0,25,50,75,100},
    axis background/.style={fill=gray!3},
    axis line style={gray!70},
    tick style={gray!70},
    grid=major,
    major grid style={line width=.35pt, draw=gray!40},
    tick label style={font=\scriptsize},
    label style={font=\small},
    title style={font=\small},
    every axis plot/.append style={line width=1.15pt},
]
\addplot+[color=figfive, mark=square*, mark options={fill=figfive!55, draw=figfive}] coordinates {(100,1.2) (200,1.6) (500,15.2) (1000,45.2) (1500,74.4)};
\addplot+[color=figten, mark=triangle*, mark options={fill=figten!60, draw=figten}] coordinates {(100,4.8) (200,27.2) (500,80.0) (1000,87.2) (1500,93.6)};
\addplot+[color=figtwentyfive, mark=diamond*, mark options={fill=figtwentyfive!60, draw=figtwentyfive}] coordinates {(100,81.2) (200,87.2) (500,99.6) (1000,100.0) (1500,100.0)};
\end{axis}
\begin{axis}[
    at={(batchtpr.south west)},
    anchor=north west,
    yshift=-0.60cm,
    width=\linewidth,
    height=0.38\linewidth,
    xlabel={Batch size},
    ylabel={FPR (\%)},
    xmin=80,
    xmax=1520,
    ymin=0,
    ymax=2,
    xtick={100,200,500,1000,1500},
    ytick={0,1,2},
    axis background/.style={fill=gray!3},
    axis line style={gray!70},
    tick style={gray!70},
    grid=major,
    major grid style={line width=.35pt, draw=gray!40},
    tick label style={font=\scriptsize},
    label style={font=\small},
    every axis plot/.append style={line width=1.15pt},
]
\addplot+[color=figfpr, mark=*, mark options={fill=figfpr, draw=figfpr}] coordinates {(100,0.4) (200,0.8) (500,0.4) (1000,1.2) (1500,0.0)};
\end{axis}
\end{tikzpicture}
\end{minipage}
\hfill
\begin{minipage}{0.49\textwidth}
\centering
\begin{tikzpicture}
\begin{axis}[
    name=thresholdtpr,
    width=\linewidth,
    height=0.54\linewidth,
    title={(b) Threshold percentile},
    ylabel={TPR (\%)},
    xmin=89.5,
    xmax=99.5,
    ymin=25,
    ymax=105,
    xtick={90,95,97.5,99},
    ytick={25,50,75,100},
    axis background/.style={fill=gray!3},
    axis line style={gray!70},
    tick style={gray!70},
    grid=major,
    major grid style={line width=.35pt, draw=gray!40},
    tick label style={font=\scriptsize},
    label style={font=\small},
    title style={font=\small},
    every axis plot/.append style={line width=1.15pt},
]
\addplot+[color=figfive, mark=square*, mark options={fill=figfive!55, draw=figfive}] coordinates {(90,83.6) (95,74.4) (97.5,63.2) (99,43.6)};
\addplot+[color=figten, mark=triangle*, mark options={fill=figten!60, draw=figten}] coordinates {(90,95.6) (95,93.6) (97.5,90.4) (99,87.6)};
\addplot+[color=figtwentyfive, mark=diamond*, mark options={fill=figtwentyfive!60, draw=figtwentyfive}] coordinates {(90,100.0) (95,100.0) (97.5,100.0) (99,100.0)};
\end{axis}
\begin{axis}[
    at={(thresholdtpr.south west)},
    anchor=north west,
    yshift=-0.60cm,
    width=\linewidth,
    height=0.38\linewidth,
    xlabel={Threshold percentile},
    ylabel={FPR (\%)},
    xmin=89.5,
    xmax=99.5,
    ymin=0,
    ymax=2,
    xtick={90,95,97.5,99},
    xticklabels={90,95,97.5,99},
    ytick={0,1,2},
    axis background/.style={fill=gray!3},
    axis line style={gray!70},
    tick style={gray!70},
    grid=major,
    major grid style={line width=.35pt, draw=gray!40},
    tick label style={font=\scriptsize},
    label style={font=\small},
    every axis plot/.append style={line width=1.15pt},
]
\addplot+[color=figfpr, mark=*, mark options={fill=figfpr, draw=figfpr}] coordinates {(90,1.6) (95,0.0) (97.5,0.0) (99,0.0)};
\end{axis}
\end{tikzpicture}
\end{minipage}

\vspace{0.2em}
\begin{tikzpicture}
\draw[color=figfive, line width=1.15pt] (0,0) -- (0.42,0);
\node[font=\scriptsize, anchor=west] at (0.48,0) {5\% TPR};
\draw[color=figten, line width=1.15pt] (1.65,0) -- (2.07,0);
\node[font=\scriptsize, anchor=west] at (2.13,0) {10\% TPR};
\draw[color=figtwentyfive, line width=1.15pt] (3.45,0) -- (3.87,0);
\node[font=\scriptsize, anchor=west] at (3.93,0) {25\% TPR};
\draw[color=figfpr, line width=1.15pt] (5.35,0) -- (5.77,0);
\node[font=\scriptsize, anchor=west] at (5.83,0) {FPR};
\end{tikzpicture}
\caption{Sensitivity analysis of the MMD detector on five representative attacker-normal pairs.
The upper panels report TPR under low attacker fractions, and the lower panels report benign FPR on a separate scale.
The 50\% and 100\% attacker-fraction TPRs remain near 100\% across the tested settings and are omitted for clarity.
Larger traffic windows improve low-fraction detection, while the 95th percentile threshold provides a strong trade-off between low benign FPR and low-fraction attacker detection.}
\label{fig:sensitivity_analysis}
\end{figure*}
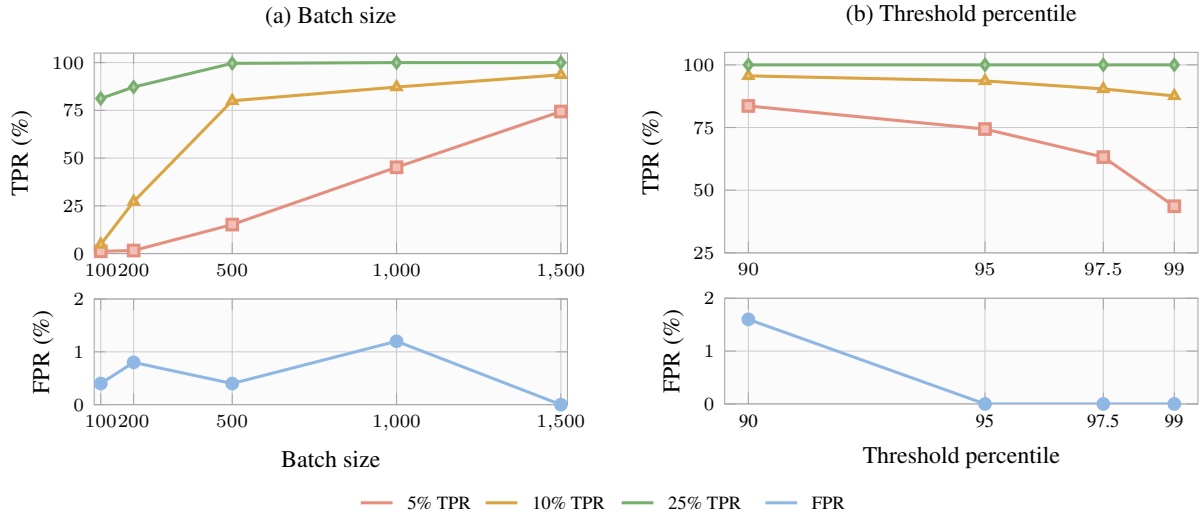

From Figure~\ref{fig:mixed_traffic_tradeoff}, we make the following observations.
(1) MMD keeps 99.7\% specificity while reaching 59.0\% TPR at 5\% attacker traffic, 93.7\% at 10\%, and 100.0\% at 25\% and 50\%.
(2) The 5\% setting remains hardest because benign queries dominate the window and the attack contributes only a weak aggregate shift.
This is especially difficult when extraction queries are semantically close to benign task inputs, such as Wikipedia-derived sentences, SQuAD-style questions, or other natural task queries.
(3) Compared with adapted baselines, MMD gives a stronger mixed-traffic trade-off.
Mahalanobis reaches 70.5\% TPR at 5\% but with lower specificity, while pseudo-energy, DATE, SEAT, CAP, and PRADA obtain 54.0\%, 39.7\%, 31.7\%, 26.4\%, and 18.9\%.
At 10\%, MMD reaches 93.7\%, outperforming all baselines.
Thus, MMD remains useful beyond pure attacker-user detection when extraction queries are diluted through cover queries, slow extraction, or distributed multi-account querying over time.

\subsection{Adaptive Evasion Analysis}

To answer RQ4, we evaluate whether an attacker can reduce the MMD signal while preserving extraction quality using MeaeQ/SST-2 as a representative setting.
We compare the original query set with two adaptive variants.
The regularized variant selects queries with an MMD penalty, while the projection variant selects queries closer to the benign acceptance region.
Table~\ref{tab:adaptive_meaeq} reports mixed-traffic detection and the resulting student's accuracy and agreement with the victim model on the held-out SST-2 task.

From Table~\ref{tab:adaptive_meaeq}, we make the following observations.
(1) Adaptive query selection reduces detection in the hardest low-fraction regime: 5\% TPR decreases from 84.0\% for original MeaeQ to 70.0\% for the regularized variant and 78.0\% for the projection variant.
(2) Detection remains strong when attacker traffic reaches 10\% or more of the window, where all three variants obtain 100.0\% TPR.
(3) Reduced detectability comes with lower extraction quality.
Student accuracy drops from 76.38\% to 67.43\% and 67.09\%, while agreement drops from 78.33\% to 70.07\% and 70.87\%.
Thus, adaptive pressure can reduce low-fraction detection, but in this setting it also degrades the extracted student's downstream utility as a cloned model.

\begin{table}[t]
\centering
\small
\setlength{\tabcolsep}{0pt}
\renewcommand{\arraystretch}{1.05}
\begin{tabular*}{\columnwidth}{@{\extracolsep{\fill}}lcccc}
\toprule
Attack & 5\% & $\geq$10\% & Acc. & Agree. \\
\midrule
Original MeaeQ & 84.0 & 100.0 & 76.38 & 78.33 \\
Regularized & 70.0 & 100.0 & 67.43 & 70.07 \\
Projection & 78.0 & 100.0 & 67.09 & 70.87 \\
\bottomrule
\end{tabular*}
\caption{Adaptive MeaeQ evaluation on SST-2.
The 5\% and $\geq$10\% columns report MMD TPR; Acc. and Agree. report student accuracy and student-victim agreement.}
\label{tab:adaptive_meaeq}
\end{table}

\begin{table}[t]
\centering
\small
\setlength{\tabcolsep}{0pt}
\renewcommand{\arraystretch}{1.05}
\begin{tabular*}{\columnwidth}{@{\extracolsep{\fill}}lcccc}
\toprule
Contam. & FPR & 100\% TPR & 5\% TPR & 10\% TPR \\
\midrule
0\% & 1.1 & 100.0 & 59.7 & 90.0 \\
1\% & 1.1 & 100.0 & 31.0 & 88.6 \\
5\% & 62.7 & 100.0 & 0.3 & 57.4 \\
10\% & 93.9 & 100.0 & 62.6 & 0.0 \\
\bottomrule
\end{tabular*}
\caption{Effect of attacker contamination in the benign reference pool.
Values are averaged over fourteen attacker-normal pairs at seed 42.}
\label{tab:reference_contamination}
\end{table}

\subsection{Reference Contamination Analysis}

To answer RQ5, we test the clean-reference assumption by injecting attacker queries into the benign reference pool at 1\%, 5\%, and 10\%, then recomputing the MMD threshold and evaluating held-out benign, pure-attacker, 5\%, and 10\% mixed traffic windows.

From Table~\ref{tab:reference_contamination}, we make the following observations.
(1) Light contamination mainly weakens the hardest low-fraction setting.
At 1\% contamination, benign FPR remains 1.1\%, pure-attacker TPR remains 100.0\%, and 10\% mixed-traffic TPR remains 88.6\%, while 5\% TPR drops from 59.7\% to 31.0\%.
(2) Sustained high contamination is a failure mode for benign-reference calibration.
At 5\% and 10\% contamination, the reference distribution shifts toward a benign-attacker mixture, causing clean benign windows to appear anomalous and making mixed windows with similar attacker fractions harder to distinguish.
(3) These results indicate that reference-pool hygiene is important in deployment.
Trusted seed references, controlled reference updates, rolling validation, and robust reference estimation can reduce the risk that slow extraction traffic enters the calibration pool.
Thus, benign-only calibration is stable under light contamination for stronger attack windows, but substantial reference poisoning can undermine both FPR control and low-fraction detection.

\subsection{Original Protocol Transfer}
\label{sec:original_baseline_protocols}

To answer RQ6, we test whether original baseline protocols transfer directly to semantic extraction-query traffic.
For each baseline with a directly identifiable original protocol, we preserve its original scoring and decision unit when possible; Appendix~\ref{app:original_baseline_protocols} provides the metric-aligned details for this protocol comparison.

From these original-protocol results, we make the following observations.
(1) PRADA does not transfer cleanly from image-query stealing detection to semantic text-query traffic: it raises alarms on 51.6\% of benign-user streams while detecting only 49.0\% of attacker-user streams.
This suggests that nearest-neighbor distance normality is brittle after moving from image queries to semantic text-query embeddings.
(2) SEAT also loses its original account-level reliability, giving 14.3\% benign FPR with only 15.0\% attacker-user TPR.
Its visual similar-pair structure does not directly carry over to NLP extraction queries.
(3) DATE and Mahalanobis lose much of their original anomaly-detection strength, dropping to 49.5 and 70.1 AUROC on individual extraction queries, far below the high AUROC values reported in their original anomaly or OOD settings.
This indicates that model extraction queries are not generic text anomalies or ordinary OOD utterances.
(4) CAP is not directly comparable as an original passive detector because its original protocol is an active output-perturbation defense; used passively, its coverage score gives 1.9\% FPR but only 21.4\% TPR.
These failures show that original protocols do not transfer directly to semantic extraction-query traffic, justifying the unified adaptations used in Section~4.3 in our main evaluation.

\subsection{Sensitivity Analysis}
\label{sec:sensitivity_decision_analysis}

To answer RQ7, we test sensitivity to the sentence encoder, MMD traffic-window size, threshold percentile, and one-sided versus two-sided baseline decisions.
Table~\ref{tab:encoder_sensitivity}, Figure~\ref{fig:sensitivity_analysis}, and Table~\ref{tab:decision_direction} summarize these results.
Appendix~\ref{app:additional_sensitivity_runtime} provides additional parameter and runtime sensitivity results for reproducibility checks.

\begin{table*}[!t]
\centering
\small
\setlength{\tabcolsep}{0pt}
\begin{tabular*}{\textwidth}{@{\extracolsep{\fill}}lccccc}
\toprule
\textbf{Encoder} & \textbf{Benign FPR} & \textbf{5\% TPR} & \textbf{10\% TPR} & \textbf{Avg. TPR} & \textbf{Balanced Acc.} \\
\midrule
\textbf{\texttt{bge-small-en-v1.5}} & \textbf{0.3} & 60.9 & 91.4 & 90.5 & 95.1 \\
\texttt{all-MiniLM-L6-v2} & \textbf{0.3} & 52.0 & 97.0 & 89.8 & 94.8 \\
\texttt{all-mpnet-base-v2} & \textbf{0.3} & 64.4 & \textbf{100.0} & 92.9 & 96.3 \\
\texttt{e5-small-v2} & 1.7 & 71.0 & 96.7 & 93.5 & 95.9 \\
\texttt{e5-base-v2} & 0.6 & \textbf{72.1} & 98.1 & \textbf{94.1} & \textbf{96.7} \\
\bottomrule
\end{tabular*}
\caption{MMD sensitivity to sentence embedding model.
All values are percentages and are averaged over the fourteen attacker-normal pairs at batch size 1,500 in the encoder-sensitivity setting.}
\label{tab:encoder_sensitivity}
\end{table*}

\begin{table*}[!t]
\centering
\scriptsize
\setlength{\tabcolsep}{3pt}
\resizebox{\textwidth}{!}{%
\begin{tabular}{lrrrrrrr|rrrrrrr|r}
\toprule
\textbf{Method} & \multicolumn{7}{c}{\textbf{One-sided}} & \multicolumn{7}{c}{\textbf{Two-sided}} & \cellcolor{gaincol}\textbf{$\Delta$ Avg.} \\
\cmidrule(lr){2-8} \cmidrule(lr){9-15} \cmidrule(lr){16-16}
 & \textbf{FPR} & \textbf{5\%} & \textbf{10\%} & \textbf{25\%} & \textbf{50\%} & \textbf{100\%} & \textbf{Avg.} & \textbf{FPR} & \textbf{5\%} & \textbf{10\%} & \textbf{25\%} & \textbf{50\%} & \textbf{100\%} & \textbf{Avg.} & \cellcolor{gaincol}\textbf{Gain} \\
\midrule
PRADA & 4.1 & 13.0 & 15.4 & 19.7 & 27.0 & 29.1 & 20.9 & 8.6 & 21.9 & 29.0 & 61.1 & 66.7 & 78.0 & \textbf{51.3} & \cellcolor{gaincol}\textbf{+30.4} \\
SEAT & 7.0 & 9.1 & 21.6 & 34.9 & 39.7 & 64.3 & 33.9 & 13.0 & 33.9 & 66.7 & 93.3 & 97.0 & 100.0 & \textbf{78.2} & \cellcolor{gaincol}\textbf{+44.3} \\
CAP & 6.6 & 18.9 & 23.3 & 27.0 & 28.6 & 28.6 & 25.3 & 12.6 & 28.3 & 41.1 & 65.7 & 88.4 & 100.0 & \textbf{64.7} & \cellcolor{gaincol}\textbf{+39.4} \\
DATE & 4.0 & 36.4 & 47.7 & 58.4 & 61.9 & 64.3 & 53.7 & 4.0 & 36.4 & 62.4 & 91.3 & 94.1 & 98.0 & \textbf{76.5} & \cellcolor{gaincol}\textbf{+22.8} \\
\bottomrule
\end{tabular}
}
\caption{Effect of one-sided and two-sided decision rules for adapted baselines on the fourteen attacker-normal query pairs.
All values are percentages.
Avg. TPR averages over attacker fractions $\rho\in\{0.05,0.10,0.25,0.50,1.00\}$.
$\Delta$ Avg. reports the absolute gain of two-sided over one-sided Avg. TPR.}
\label{tab:decision_direction}
\end{table*}

From Table~\ref{tab:encoder_sensitivity}, Figure~\ref{fig:sensitivity_analysis}, and Table~\ref{tab:decision_direction}, we make the following observations.
(1) MMD is not tied to a single sentence encoder.
Across five encoders, benign FPR remains 0.3--1.7\%, Avg. TPR remains 89.8--94.1\%, and Balanced Acc. remains 94.8--96.7\%.
The 5\% setting is more encoder-sensitive, ranging from 52.0\% to 72.1\% TPR.
(2) Larger traffic windows substantially improve low-fraction attacker detection while keeping benign FPR low.
At 5\% attacker traffic, TPR increases from 1.2\% with 100-query windows to 74.4\% with 1,500-query windows; at 10\%, it increases from 4.8\% to 93.6\%, while FPR remains between 0.0\% and 1.2\%.
(3) The threshold percentile controls the trade-off between low-FPR operation and early detection of highly diluted attacks.
The 95th percentile is a deliberately conservative default: it keeps FPR at 0.0\%, while retaining 74.4\% TPR at 5\% attacker traffic and 93.6\% TPR at 10\%.
Lowering the threshold to the 90th percentile improves 5\% TPR from 74.4\% to 83.6\%, but increases benign FPR from 0.0\% to 1.6\%.
(4) Stronger attacker fractions are stable across these choices.
For 25\%, 50\%, and pure attacker traffic, MMD remains close to or at 100.0\% TPR across the tested batch sizes and threshold percentiles.
(5) Two-sided decision rules are important for adapted baselines.
They improve Avg. TPR for PRADA, SEAT, CAP, and DATE by 30.4, 44.3, 39.4, and 22.8 points, respectively, showing that text-query extraction traffic can deviate from benign calibration in either direction.
Overall, the default encoder is strong but not uniquely effective, the chosen window size and threshold provide a practical low-FPR operating point, and two-sided baseline decisions avoid brittle directional assumptions.
Deployments prioritizing early detection can use a less conservative threshold, but highly diluted and semantically aligned extraction traffic remains challenging.
Taken together, these sensitivity results clarify how the detector should be used in practice.
The main configuration is conservative: it favors low benign FPR and reliable detection once extraction traffic is no longer extremely diluted.
Operators with different alert budgets can adjust the threshold or window size, but the resulting trade-off should be reported together with benign FPR because small changes in early-detection sensitivity can substantially affect the number of benign alerts.

\section{Conclusion}

In this paper, we introduced a benign-calibrated traffic-window formulation for monitoring model extraction attacks in LLM API traffic.
The formulation addresses two practical challenges: benign-looking individual extraction queries and diluted attack behavior in mixed normal traffic.
We instantiated it with semantic query embeddings and an MMD two-sample statistic calibrated only from benign historical queries.
Across fourteen attacker-normal query pairs, the detector is effective in both user-level and mixed multi-user traffic settings.
Compared with adapted baselines, MMD gives the best deployment-oriented trade-off, combining high detection, near-zero benign false positives, and the highest balanced accuracy.
These results suggest that simple statistical tests, when paired with benign calibration and traffic-window evaluation, can serve as strong baselines for LLM security monitoring.
More broadly, our findings support a shift from isolated query scoring toward deployment-oriented traffic-window evaluation.
Under this view, the main question is not whether every extraction query is individually anomalous, but whether repeated query behavior creates a measurable aggregate shift relative to trusted benign traffic.
This framing makes evaluation assumptions explicit: detectors should share the same query-only representation, benign-calibrated thresholding rule, and mixed-traffic windows.
This protocol can make future defenses easier to compare by separating original-setting attack detection from passive, low-FPR monitoring for deployed LLM API services.

\section{Acknowledgments}
This material is supported by the National Science Foundation (NSF) under grant \#2530786.

\section*{Limitations}

While the proposed benign-calibrated detector demonstrates strong performance across fourteen attacker-normal query pairs, we acknowledge several limitations of the current study.
First, low-fraction mixed traffic remains challenging.
The detector performs strongly once attacker queries occupy 10\% or more of a traffic window, but the 5\% setting is substantially harder, especially when extraction queries are semantically close to benign task inputs.
Deployments that prioritize earlier detection can use a less conservative threshold, but this increases benign alerts, and highly diluted extraction remains difficult.
Second, our adaptive evaluation is representative rather than exhaustive.
We test regularized and projection-style MeaeQ variants on SST-2, but stronger adaptive attackers could explicitly optimize larger query pools, use paraphrasing or style transfer, or preserve task utility while matching the benign embedding distribution more closely.
Our results should therefore be interpreted as an attack-label-free monitoring baseline rather than a complete defense against fully adaptive white-box extraction.
Third, benign reference quality and temporal stability are critical.
Our contamination analysis shows that light contamination mainly harms the hardest 5\% mixed setting, while sustained 5--10\% reference contamination can sharply increase FPR or hide mixed windows with matching attacker fractions.
In addition, benign traffic may drift over time, making a static reference pool or threshold stale.
Deployments should therefore use trusted seed references, controlled updates, rolling benign-vs-benign recalibration, drift monitoring, recent holdout validation, and robust reference estimation.

Beyond these detection-specific limitations, our evaluation scope is also bounded.
The detector uses only query text and is evaluated in an offline traffic-window setting.
Real deployments can additionally use timestamps, account metadata, response information, rate limits, and cross-window accumulation.
Although our evaluation covers four extraction scenarios and fourteen attacker-normal pairs, it does not exhaust all domains, languages, multi-turn interactions, tool-use queries, or production API traffic patterns.
Future work could address these limitations by evaluating stronger adaptive attacks, extending the detector to temporal and account-aware monitoring, and testing the framework on broader production-style API traffic.


\bibliography{anthology}

\appendix

\section{Related Work}

\paragraph{Model extraction attacks.}
Model extraction attacks aim to replicate the functionality or knowledge of a target model through black-box API access.
Early work showed that prediction APIs leak enough information for an adversary to train substitute models, and later studies extended this threat to large-scale functionality stealing, knowledge distillation from black-box services, and transfer-based adversarial attacks \citep{tramer2016stealing,orekondy2019knockoff,jagielski2020high,chandrasekaran2020exploring,hinton2015distilling,papernot2017practical}.
Recent surveys further organize these risks across LLMs and graph-based machine-learning-as-a-service IP protection \citep{zhao2025surveyllmextraction,zhao2025systematicsurvey,li2025graphip}.
In NLP, extraction attacks query deployed APIs with synthetic or natural text sources, including random words, Wikipedia-derived text, SQuAD-style prompts, task-relevant selections, and domain knowledge questions \citep{krishna2020thieves,birch2023modelleeching,dai2023meaeq,li2026queryefficient}.
Although these attacks differ in query generation strategy and victim task, they share the operational requirement of sending many individually plausible queries to a deployed model.
This makes extraction a traffic-monitoring problem: the suspicious signal may be weak at the single-query level but accumulate over a window of API requests.

\paragraph{Detection and defense against model extraction.}
Existing defenses against model extraction monitor query behavior, perturb outputs, use ensemble or distillation-based protection, or attempt to increase the adversary's query cost.
PRADA measures distributional irregularities in sequences of API queries, SEAT detects suspicious accounts with a learned similarity encoder, CAP protects prompt-tuned LLMs through coverage-aware output perturbation, and MISLEADER uses ensembles of distilled models to reduce extractability \citep{juuti2019prada,zhang2021seat,kulkarni2026stealing,cheng2025misleader}.
Related watermarking, fingerprinting, and certified ownership verification methods provide post-hoc evidence of copying rather than an online alarm during extraction \citep{adi2018turning,szyller2021dawn,shen2026credit}.
These methods address important parts of the defense problem, but their original assumptions differ from ours.
PRADA and SEAT are closest to user-level detection, where a benign account contains only legitimate queries and an attacker account contains a complete extraction workflow.
CAP is an active output-perturbation defense rather than a passive detector for incoming query traffic.
Our setting instead treats the monitored object as an incoming traffic window and calibrates detection from benign historical windows.
This captures both pure attacker traffic and mixed, potentially cross-user traffic where extraction queries are diluted by benign requests.
Our original-protocol analysis further shows that existing defense assumptions do not transfer directly to semantic text query traffic, motivating query-only adaptations under a common deployment protocol.

\paragraph{Distributional detection in embedding space.}
Distributional testing provides a natural framework for detecting query streams whose aggregate behavior deviates from benign traffic.
The maximum mean discrepancy (MMD) is a kernel two-sample test for nonparametric distribution comparison, and related criteria have been widely used for model criticism and distributional analysis \citep{gretton2007kernel,gretton2012kernel,sutherland2016generative}.
Other geometry-based or anomaly-based detectors, including Mahalanobis distance, OOD scoring, one-class objectives, and DATE, score whether individual inputs fall outside a learned in-domain representation or textual normality model \citep{lee2018simple,podolskiy2021mahalanobis,scholkopf2001estimating,hendrycks2017baseline,ruff2018deep,liu2020energy,manolache2021date}.
These signals are useful, but they typically start from individual-input scores, whereas model extraction monitoring requires reasoning about weak shifts accumulated over a traffic window of individually plausible queries.
We use semantic sentence encoders to represent text queries \citep{reimers2019sentencebert,gao2021simcse,xiao2023cpack} and evaluate detectors after adapting them to the same benign-calibrated window protocol.
Under this protocol, the role of distributional testing is not to identify a single abnormal query, but to decide whether an incoming window departs from the benign reference distribution while maintaining a low false positive rate.

\section{Broader Impact}

Our work has several potential positive societal impacts.
By improving the detection of model extraction behavior in hosted LLM services, the proposed framework can help protect model owners from intellectual-property theft and reduce incentives for unauthorized model cloning.
Because the detector is calibrated from benign traffic and does not require labeled attack examples, it can support practical monitoring for services where new extraction strategies appear over time.
The query-only design also avoids requiring access to model parameters or response contents, which can make the detector easier to integrate into service-side monitoring pipelines.

However, we also recognize potential negative impacts.
Detection systems that monitor query traffic can raise privacy and governance concerns if deployed without data minimization, access control, or retention limits.
False positives can disrupt legitimate users, especially in high-stakes or commercial API settings.
There is also a risk that public discussion of detection signals could help adaptive attackers design queries that better evade monitoring.
We encourage responsible deployment practices, including privacy-preserving logging, conservative thresholds, human review for enforcement decisions, and clear appeal mechanisms for affected users.

\section{Implementation Details}
\label{app:implementation_details}

Table~\ref{tab:dataset_sources} summarizes the attacker and normal query sources used in the fourteen attacker-normal pairs described in Section~4.1.

\begin{table*}[t]
\centering
\small
\setlength{\tabcolsep}{4pt}
\begin{tabular}{p{0.22\textwidth}p{0.38\textwidth}p{0.32\textwidth}}
\toprule
Dataset family & Attacker query source & Normal query source \\
\midrule
\textsc{Query-Efficient-Med} & Self-generated medical domain stealing queries & Medicine-related WildChat queries \\
\textsc{Model-Leeching} & Template-wrapped SQuAD extraction prompts & Original SQuAD questions \\
\textsc{MeaeQ} & MeaeQ queries generated from WikiText-103 & Hate Speech, AG News, SST-2, IMDB \\
\textsc{BERT-API} & BERT API extraction queries generated from WikiText-103 & GLUE SST-2, GLUE MNLI, SQuAD 1.1, BoolQ \\
\bottomrule
\end{tabular}
\caption{Detailed source summary for the attacker-normal query pairs described in Section~4.1.}
\label{tab:dataset_sources}
\end{table*}

\paragraph{Data splitting and batch construction.}
For each attacker-normal query pair, we split the normal queries into an 80\% benign reference pool and a 20\% benign evaluation pool.
The benign reference pool is used for calibration and for sampling benign reference batches.
The benign evaluation pool is used only for measuring benign false positives.
For each random seed, we sample 50 benign-only evaluation batches from the benign evaluation pool and 50 pure attacker batches from the attacker query pool.
For mixed-traffic evaluation, we construct 50 batches for each attacker fraction $\rho\in\{0.05,0.10,0.25,0.50\}$.
Each mixed batch contains $\lfloor \rho n \rfloor$ attacker queries and $n-\lfloor \rho n \rfloor$ benign queries, where $n$ is the traffic-window size.
Unless otherwise stated, $n=1500$.
All reported main results average over seeds 1, 20, and 42.

\paragraph{Benign calibration.}
All detectors use thresholds calibrated only from benign data.
For MMD, we construct 1,000 benign-vs-benign calibration scores by sampling pairs of benign reference batches from the benign reference pool.
The default decision threshold is the 95th percentile of these benign calibration scores.
For adapted baselines with two-sided decisions, we construct a benign interval from the lower and upper tails of the benign calibration score distribution, following the thresholding convention of each detector implementation.
A traffic window is flagged when its score falls outside the corresponding benign interval.
This keeps the score definition and decision convention of each adapted baseline unchanged while allowing deviations in either direction to be detected.
For one-sided decisions, only the original expected deviation direction is used.
Benign evaluation batches are never used for threshold calibration, so the reported FPR measures generalization from benign calibration data to held-out benign traffic.

\paragraph{MMD implementation.}
All queries are embedded with \texttt{BAAI/bge-small-en-v1.5}, and embeddings are normalized before scoring.
For a test traffic window, we sample 20 benign reference batches with the same batch size from the benign reference pool.
We compute the unbiased squared MMD between the test embeddings and each benign reference batch, then average the 20 scores to obtain the final window score.
The kernel is a multi-kernel RBF kernel.
We set the base bandwidth by the median heuristic on benign embeddings and use four bandwidths obtained by multiplying the base bandwidth by $\{0.5,1,2,4\}$.
The averaged MMD score is compared with the benign-calibrated threshold described above.

\paragraph{Adapted baseline implementation.}
For PRADA, we compute nearest-neighbor distance statistics in the shared embedding space and apply benign-calibrated two-sided thresholding.
For SEAT, we compute the ratio of similar query pairs inside each traffic window.
The pair similarity cutoff is set by the 99th percentile of benign pair similarities, and the resulting batch-level ratio is thresholded using benign calibration.
For CAP, we use stream mode with one batch per stream and compute a coverage-style score over the query embeddings.
Because the original CAP method is an active output-perturbation defense, our adapted version uses only its coverage signal as a passive traffic-window score.
For DATE, we train the self-supervised anomaly detector on benign training queries using an ELECTRA-small backbone.
We score each query by its replaced-token detection anomaly score and aggregate query scores by the batch mean.
For Mahalanobis, we fit a benign centroid and covariance model in embedding space.
We use a ridge value of $10^{-6}$ for covariance stabilization and score a traffic window by the mean marginal Mahalanobis distance of its query embeddings.
For pseudo-class energy, we cluster benign embeddings into pseudo classes, derive pseudo logits from distances to cluster centers, compute query-level energy scores, and aggregate them by the batch mean before benign-calibrated thresholding.

\section{Original Baseline Protocols}
\label{app:original_baseline_protocols}

This appendix provides the protocol bookkeeping behind Section~\ref{sec:original_baseline_protocols}.
Table~\ref{tab:original_protocol_transfer} records three pieces of information for each original-protocol baseline: the metric used by the original paper, whether the same metric can be computed on our query pairs, and our endpoint diagnostic under the transferred original protocol.
We keep the original decision unit whenever possible and evaluate only benign-user and attacker-user cases.
Mixed-traffic rows are excluded because several original protocols are not defined as passive mixed-window detectors.

\begin{table*}[t]
\centering
\small
\setlength{\tabcolsep}{3pt}
\begin{tabular}{p{0.12\textwidth}p{0.33\textwidth}p{0.27\textwidth}p{0.20\textwidth}}
\toprule
\textbf{Method} & \textbf{Original paper metric and result} & \textbf{Same metric on our data} & \textbf{Our transfer diagnostic} \\
\midrule
PRADA & Sequence-level detection and benign FPR; main text reports 100\% detection with no false positives, and Table~VIII reports 0.0--0.6\% FPR under tested thresholds \citep{juuti2019prada}. & Same metric is available: 51.6\% benign FPR and 49.0\% attacker-stream detection. & Benign FPR / attacker TPR: 51.6\% / 49.0\%. \\
SEAT & Benign-account FPR and attacker account cost; benign FPR is below 0.05\% across 13 image query sets, and adaptive attackers require 3.8--16$\times$ more accounts \citep{zhang2021seat}. & Benign FPR is available: 14.3\%. Account-cost multiplication is not available because our logs do not simulate repeated account creation until extraction succeeds. & Benign FPR / attacker TPR: 14.3\% / 15.0\%. \\
DATE & Text anomaly AUROC; DATE reports 69.7--95.9 AUROC on 20Newsgroups and AG News anomaly splits \citep{manolache2021date}. & Same metric is available: 49.5 AUROC from individual-query anomaly scores. & Benign FPR / attacker TPR: 5.0\% / 12.5\%. \\
Mahalanobis & OOD intent AUROC; RoBERTa-Mahalanobis reports 97.6--99.8 AUROC on CLINC150, ROSTD, ROSTD-Coarse, and SNIPS \citep{podolskiy2021mahalanobis}. & Same metric is available: 70.1 AUROC from individual-query Mahalanobis scores. & Benign FPR / attacker TPR: 5.1\% / 19.0\%. \\
CAP & Downstream utility under active perturbation; attacker accuracy drops from 86.00--93.60\% to 50.86--52.90\%, while legitimate-user accuracy remains 78.54--88.20\% \citep{kulkarni2026stealing}. & Same metric is not available without running downstream extraction and utility evaluation. & Benign FPR / attacker TPR: 1.9\% / 21.4\%. \\
\bottomrule
\end{tabular}
\caption{Metric-aligned transfer analysis of original baseline protocols.
The second column reports the metric and result used by the original paper.
The third column reports the same metric on our attacker-normal query pairs when it can be computed.
The final column reports our decision-level transfer diagnostic for judging whether the original protocol is usable in our setting.}
\label{tab:original_protocol_transfer}
\end{table*}

\paragraph{Evaluation unit.}
For this appendix, we use only the two endpoint cases: benign-user traffic and attacker-user traffic.
Benign-user traffic contains only legitimate queries from the held-out benign evaluation split.
Attacker-user traffic contains only extraction queries generated by the corresponding model extraction method.
Mixed-traffic windows are excluded from the original-protocol transfer table because several original baselines do not define a passive mixed-window decision unit.

\paragraph{Metric alignment.}
We separate two quantities for each original-protocol baseline.
The first is the metric reported by the original paper, such as sequence-level detection, benign-account FPR, AUROC, or downstream extraction utility.
The second is our decision diagnostic, which reports benign FPR and attacker-user TPR after applying the transferred original protocol to our query pairs.
We only compare original-paper numbers with our numbers when the metric is aligned.
When the metric is not aligned, the table marks the mismatch and uses our decision diagnostic only to assess whether the original protocol is usable in our setting.

\paragraph{Score construction.}
For PRADA, the original score is computed on query streams using nearest-neighbor distance deviations.
For SEAT, we preserve its account-level similar-pair decision unit and evaluate benign accounts and attacker accounts.
For DATE and Mahalanobis, we compute individual-query anomaly scores and report AUROC because this matches the threshold-free metrics used by the original papers.
For CAP, we transfer the coverage-style score as a passive diagnostic, but we do not reproduce its original downstream utility evaluation because that would require a full extraction-and-defense pipeline rather than stored query logs.

\paragraph{Relation to the unified protocol.}
The unified comparison in Section~4.3 uses the same semantic representation, batch construction, and benign-calibrated thresholding across all methods.
The original-protocol analysis in Section~\ref{sec:original_baseline_protocols} and Table~\ref{tab:original_protocol_transfer} deliberately avoids these adaptations.
This separation allows the main text to answer two different questions: whether original protocols transfer directly, and whether minimally adapted scoring rules provide useful baselines under a common text-query traffic protocol.

\section{Additional Sensitivity and Runtime Analysis}
\label{app:additional_sensitivity_runtime}

\paragraph{Reference repeat sensitivity.}
We next study the effect of the number of benign reference batches averaged for each incoming traffic window.
Specifically, we vary the number of reference repeats in $\{1,5,10,20\}$ while keeping the other MMD settings fixed.
We show the results in Table~\ref{tab:appendix_reference_repeats}.
From Table~\ref{tab:appendix_reference_repeats}, we make the following observations.
(1) A single reference repeat is fast but less stable, increasing benign FPR to 5.3\%.
(2) Five repeats already recover the low-FPR behavior of the default setting, with 0.3\% benign FPR, 90.3\% Avg. TPR, and 95.0\% Balanced Acc.
(3) Increasing repeats from 5 to 20 provides only marginal accuracy changes while increasing running time.
In conclusion, the default value of 20 is conservative, but a lightweight configuration with 5 repeats offers a similar detection trade-off at lower cost.

\begin{table*}[t]
\centering
\small
\setlength{\tabcolsep}{0pt}
\begin{tabular*}{\textwidth}{@{\extracolsep{\fill}}rrrrrrr}
\toprule
\textbf{Repeats} & \textbf{Seconds} & \textbf{Sec./Unit} & \textbf{Benign FPR} & \textbf{5\% TPR} & \textbf{Avg. TPR} & \textbf{Balanced Acc.} \\
\midrule
1  & \textbf{45.2} & \textbf{0.151} & 5.3 & 56.4 & 89.6 & 92.2 \\
\textbf{5}  & 55.0 & 0.183 & \textbf{0.3} & 59.4 & 90.3 & 95.0 \\
10 & 68.1 & 0.227 & 1.1 & 59.6 & 90.1 & 94.5 \\
\textbf{20} & 94.1 & 0.314 & \textbf{0.3} & \textbf{60.9} & \textbf{90.5} & \textbf{95.1} \\
\bottomrule
\end{tabular*}
\caption{MMD sensitivity to the number of benign reference repeats.
Seconds report the mean wall-clock time over the fourteen attacker-normal pairs; detection values are percentages.}
\label{tab:appendix_reference_repeats}
\end{table*}

\paragraph{Null sample sensitivity.}
We also vary the number of benign-vs-benign null samples used to estimate the calibration threshold.
Specifically, we test $\{100,250,500,1000\}$ null samples while keeping the reference repeat count fixed at 20.
We show the results in Table~\ref{tab:appendix_null_samples}.
From Table~\ref{tab:appendix_null_samples}, we make the following observations.
(1) Detection performance is stable across a wide range of null sample counts.
All settings obtain about 90\% Avg. TPR and about 95\% Balanced Acc.
(2) Using 250 or 500 null samples slightly improves 5\% TPR over the default 1,000-sample setting, while preserving low benign FPR.
(3) The runtime difference is modest because the main cost is still embedding and repeated MMD scoring.
In conclusion, 1,000 null samples is a conservative default, while 250 null samples is a reasonable lightweight alternative.

\begin{table*}[t]
\centering
\small
\setlength{\tabcolsep}{0pt}
\begin{tabular*}{\textwidth}{@{\extracolsep{\fill}}rrrrrrr}
\toprule
\textbf{Null Samples} & \textbf{Seconds} & \textbf{Sec./Unit} & \textbf{Benign FPR} & \textbf{5\% TPR} & \textbf{Avg. TPR} & \textbf{Balanced Acc.} \\
\midrule
100  & \textbf{85.5} & \textbf{0.285} & \textbf{0.0} & 57.6 & 90.0 & 95.0 \\
\textbf{250}  & 87.1 & 0.290 & 0.6 & 62.9 & \textbf{90.9} & \textbf{95.1} \\
500  & 89.1 & 0.297 & 0.6 & \textbf{63.4} & 90.8 & \textbf{95.1} \\
\textbf{1000} & 93.0 & 0.310 & 0.3 & 60.9 & 90.5 & \textbf{95.1} \\
\bottomrule
\end{tabular*}
\caption{MMD sensitivity to the number of null samples used for benign calibration.
Seconds report the mean wall-clock time over the fourteen attacker-normal pairs; detection values are percentages.}
\label{tab:appendix_null_samples}
\end{table*}

\paragraph{End-to-end wall-clock runtime.}
Finally, we compare end-to-end wall-clock running time on \textsc{Model-Leeching} with batch size 1,500 and seed 42.
The timing includes the full existing run scripts and 300 evaluation units.
Table~\ref{tab:appendix_runtime} reports the default MMD configuration, where MMD uses 20 reference repeats and 1,000 null samples, a lightweight MMD configuration with 5 reference repeats and 250 null samples, and one run for each adapted baseline.
From Table~\ref{tab:appendix_runtime}, we make the following observations.
(1) Under the default configuration, MMD takes 395.7 seconds, comparable to CAP and moderately slower than PRADA, Mahalanobis, and SEAT.
(2) The lightweight MMD configuration reduces MMD runtime to 350.6 seconds, an 11.4\% reduction, while the parameter sensitivity results above show little loss in detection performance.
(3) Pseudo-energy is the fastest method in this timing sweep, while DATE is substantially slower than the other methods because the measured run includes DATE training.
In conclusion, MMD is not only effective but also practical: its full configuration is comparable to several adapted baselines, and its lightweight configuration further reduces cost.

\begin{table*}[t]
\centering
\small
\setlength{\tabcolsep}{0pt}
\begin{tabular*}{\textwidth}{@{\extracolsep{\fill}}lrr}
\toprule
\textbf{Method} & \textbf{Seconds} & \textbf{Sec./Unit} \\
\midrule
MMD-default & 395.7 & 1.319 \\
MMD-lightweight & 350.6 & 1.169 \\
Mahalanobis & 339.1 & 1.130 \\
PRADA & 336.0 & 1.120 \\
SEAT & 273.0 & 0.910 \\
Pseudo-energy & \textbf{53.3} & \textbf{0.178} \\
CAP & 393.8 & 1.313 \\
DATE & 857.6 & 2.859 \\
\bottomrule
\end{tabular*}
\caption{End-to-end cold-cache wall-clock runtime on \textsc{Model-Leeching} with batch size 1,500 and seed 42.
MMD-default uses 20 reference repeats and 1,000 null samples; MMD-lightweight uses 5 reference repeats and 250 null samples.
DATE includes training time.}
\label{tab:appendix_runtime}
\end{table*}

\section{Detailed Per-Dataset Results}
\label{app:per_dataset_results}

Table~\ref{tab:main_results} reports aggregate performance over all attacker-normal pairs.
This appendix expands the main results by reporting one table for each attacker-normal pair.
Each table uses the same unified traffic-window protocol as Table~\ref{tab:main_results}.
Benign FPR is computed on benign-only evaluation batches, and the TPR columns are computed on mixed or pure-attacker evaluation batches.
Avg. TPR averages over attacker fractions $\rho\in\{0.05,0.10,0.25,0.50,1.00\}$.
Balanced Acc. is computed as $(\mathrm{Avg.\ TPR}+100-\mathrm{FPR})/2$.
All values are percentages and are reported as mean $\pm$ standard deviation over three random seeds.

\begin{table*}[t]
\centering
\footnotesize
\setlength{\tabcolsep}{2.2pt}
\begin{tabular}{lcccccccc}
\toprule
Method & Benign FPR & 5\% TPR & 10\% TPR & 25\% TPR & 50\% TPR & 100\% TPR & Avg. TPR & Balanced Acc. \\
\midrule
MMD & 0.0 $\pm$ 0.0 & 46.7 $\pm$ 37.0 & 100.0 $\pm$ 0.0 & 100.0 $\pm$ 0.0 & 100.0 $\pm$ 0.0 & 100.0 $\pm$ 0.0 & 89.3 $\pm$ 7.4 & 94.7 $\pm$ 3.7 \\
Mahalanobis & 33.3 $\pm$ 47.1 & 33.3 $\pm$ 47.1 & 85.3 $\pm$ 14.3 & 100.0 $\pm$ 0.0 & 100.0 $\pm$ 0.0 & 100.0 $\pm$ 0.0 & 83.7 $\pm$ 11.7 & 75.2 $\pm$ 17.8 \\
DATE & 22.7 $\pm$ 22.9 & 24.0 $\pm$ 19.6 & 60.7 $\pm$ 39.0 & 100.0 $\pm$ 0.0 & 100.0 $\pm$ 0.0 & 100.0 $\pm$ 0.0 & 76.9 $\pm$ 11.2 & 77.1 $\pm$ 16.8 \\
SEAT & 0.7 $\pm$ 0.9 & 22.0 $\pm$ 19.3 & 59.3 $\pm$ 26.2 & 100.0 $\pm$ 0.0 & 100.0 $\pm$ 0.0 & 100.0 $\pm$ 0.0 & 76.3 $\pm$ 9.0 & 87.8 $\pm$ 4.1 \\
Pseudo-energy & 11.3 $\pm$ 16.2 & 30.7 $\pm$ 21.0 & 38.0 $\pm$ 23.6 & 70.7 $\pm$ 15.0 & 99.3 $\pm$ 1.2 & 100.0 $\pm$ 0.0 & 67.7 $\pm$ 11.9 & 78.2 $\pm$ 3.0 \\
CAP & 40.0 $\pm$ 22.9 & 12.7 $\pm$ 6.2 & 2.0 $\pm$ 1.6 & 30.7 $\pm$ 20.4 & 92.0 $\pm$ 11.3 & 100.0 $\pm$ 0.0 & 47.5 $\pm$ 1.9 & 53.7 $\pm$ 12.0 \\
PRADA & 2.7 $\pm$ 3.8 & 16.7 $\pm$ 11.1 & 7.3 $\pm$ 5.7 & 34.0 $\pm$ 2.8 & 100.0 $\pm$ 0.0 & 100.0 $\pm$ 0.0 & 51.6 $\pm$ 2.6 & 74.5 $\pm$ 0.6 \\
\bottomrule
\end{tabular}
\caption{Per-method detection results on Query-Efficient-Med.}
\label{tab:per_dataset_query_efficent_med}
\end{table*}

\begin{table*}[t]
\centering
\footnotesize
\setlength{\tabcolsep}{2.2pt}
\begin{tabular}{lcccccccc}
\toprule
Method & Benign FPR & 5\% TPR & 10\% TPR & 25\% TPR & 50\% TPR & 100\% TPR & Avg. TPR & Balanced Acc. \\
\midrule
MMD & 0.0 $\pm$ 0.0 & 100.0 $\pm$ 0.0 & 100.0 $\pm$ 0.0 & 100.0 $\pm$ 0.0 & 100.0 $\pm$ 0.0 & 100.0 $\pm$ 0.0 & 100.0 $\pm$ 0.0 & 100.0 $\pm$ 0.0 \\
Mahalanobis & 36.0 $\pm$ 45.3 & 50.0 $\pm$ 35.6 & 85.3 $\pm$ 10.5 & 100.0 $\pm$ 0.0 & 100.0 $\pm$ 0.0 & 100.0 $\pm$ 0.0 & 87.1 $\pm$ 9.2 & 75.5 $\pm$ 18.1 \\
DATE & 32.0 $\pm$ 45.3 & 92.0 $\pm$ 11.3 & 100.0 $\pm$ 0.0 & 100.0 $\pm$ 0.0 & 100.0 $\pm$ 0.0 & 100.0 $\pm$ 0.0 & 98.4 $\pm$ 2.3 & 83.2 $\pm$ 22.1 \\
SEAT & 6.0 $\pm$ 7.1 & 100.0 $\pm$ 0.0 & 100.0 $\pm$ 0.0 & 100.0 $\pm$ 0.0 & 100.0 $\pm$ 0.0 & 100.0 $\pm$ 0.0 & 100.0 $\pm$ 0.0 & 97.0 $\pm$ 3.6 \\
Pseudo-energy & 1.3 $\pm$ 1.2 & 100.0 $\pm$ 0.0 & 100.0 $\pm$ 0.0 & 100.0 $\pm$ 0.0 & 100.0 $\pm$ 0.0 & 100.0 $\pm$ 0.0 & 100.0 $\pm$ 0.0 & 99.3 $\pm$ 0.6 \\
CAP & 18.7 $\pm$ 25.0 & 32.0 $\pm$ 41.0 & 36.7 $\pm$ 42.3 & 93.3 $\pm$ 9.4 & 100.0 $\pm$ 0.0 & 100.0 $\pm$ 0.0 & 72.4 $\pm$ 17.7 & 76.9 $\pm$ 3.8 \\
PRADA & 2.7 $\pm$ 2.5 & 2.0 $\pm$ 1.6 & 0.0 $\pm$ 0.0 & 92.0 $\pm$ 1.6 & 56.7 $\pm$ 9.3 & 94.7 $\pm$ 2.5 & 49.1 $\pm$ 2.5 & 73.2 $\pm$ 1.8 \\
\bottomrule
\end{tabular}
\caption{Per-method detection results on Model-Leeching.}
\label{tab:per_dataset_model_leeching}
\end{table*}

\begin{table*}[t]
\centering
\footnotesize
\setlength{\tabcolsep}{2.2pt}
\begin{tabular}{lcccccccc}
\toprule
Method & Benign FPR & 5\% TPR & 10\% TPR & 25\% TPR & 50\% TPR & 100\% TPR & Avg. TPR & Balanced Acc. \\
\midrule
MMD & 0.0 $\pm$ 0.0 & 54.0 $\pm$ 21.6 & 100.0 $\pm$ 0.0 & 100.0 $\pm$ 0.0 & 100.0 $\pm$ 0.0 & 100.0 $\pm$ 0.0 & 90.8 $\pm$ 4.3 & 95.4 $\pm$ 2.2 \\
Mahalanobis & 0.0 $\pm$ 0.0 & 99.3 $\pm$ 0.9 & 100.0 $\pm$ 0.0 & 100.0 $\pm$ 0.0 & 100.0 $\pm$ 0.0 & 100.0 $\pm$ 0.0 & 99.9 $\pm$ 0.2 & 99.9 $\pm$ 0.1 \\
DATE & 31.3 $\pm$ 27.0 & 19.3 $\pm$ 15.7 & 32.0 $\pm$ 22.1 & 70.0 $\pm$ 21.2 & 85.3 $\pm$ 20.7 & 100.0 $\pm$ 0.0 & 61.3 $\pm$ 4.9 & 65.0 $\pm$ 16.0 \\
SEAT & 5.3 $\pm$ 2.5 & 46.7 $\pm$ 32.1 & 79.3 $\pm$ 29.2 & 100.0 $\pm$ 0.0 & 100.0 $\pm$ 0.0 & 100.0 $\pm$ 0.0 & 85.2 $\pm$ 12.2 & 89.9 $\pm$ 7.0 \\
Pseudo-energy & 0.0 $\pm$ 0.0 & 41.3 $\pm$ 44.1 & 99.3 $\pm$ 1.2 & 100.0 $\pm$ 0.0 & 100.0 $\pm$ 0.0 & 100.0 $\pm$ 0.0 & 88.1 $\pm$ 8.9 & 94.1 $\pm$ 4.5 \\
CAP & 12.0 $\pm$ 15.6 & 42.0 $\pm$ 22.7 & 64.7 $\pm$ 21.7 & 90.7 $\pm$ 11.8 & 99.3 $\pm$ 0.9 & 100.0 $\pm$ 0.0 & 79.3 $\pm$ 10.1 & 83.7 $\pm$ 5.3 \\
PRADA & 5.3 $\pm$ 4.1 & 12.0 $\pm$ 9.1 & 46.0 $\pm$ 24.7 & 100.0 $\pm$ 0.0 & 100.0 $\pm$ 0.0 & 100.0 $\pm$ 0.0 & 71.6 $\pm$ 6.6 & 83.1 $\pm$ 4.8 \\
\bottomrule
\end{tabular}
\caption{Per-method detection results on MeaeQ-HateSpeech.}
\label{tab:per_dataset_meaeq_hatespeech}
\end{table*}

\begin{table*}[t]
\centering
\footnotesize
\setlength{\tabcolsep}{2.2pt}
\begin{tabular}{lcccccccc}
\toprule
Method & Benign FPR & 5\% TPR & 10\% TPR & 25\% TPR & 50\% TPR & 100\% TPR & Avg. TPR & Balanced Acc. \\
\midrule
MMD & 0.0 $\pm$ 0.0 & 28.0 $\pm$ 27.9 & 99.3 $\pm$ 0.9 & 100.0 $\pm$ 0.0 & 100.0 $\pm$ 0.0 & 100.0 $\pm$ 0.0 & 85.5 $\pm$ 5.7 & 92.7 $\pm$ 2.9 \\
Mahalanobis & 0.7 $\pm$ 0.9 & 98.7 $\pm$ 0.9 & 100.0 $\pm$ 0.0 & 100.0 $\pm$ 0.0 & 100.0 $\pm$ 0.0 & 100.0 $\pm$ 0.0 & 99.7 $\pm$ 0.2 & 99.5 $\pm$ 0.4 \\
DATE & 0.7 $\pm$ 0.9 & 16.0 $\pm$ 19.9 & 42.7 $\pm$ 36.2 & 96.7 $\pm$ 4.7 & 100.0 $\pm$ 0.0 & 100.0 $\pm$ 0.0 & 71.1 $\pm$ 11.8 & 85.2 $\pm$ 6.3 \\
SEAT & 44.0 $\pm$ 30.2 & 38.7 $\pm$ 43.6 & 72.0 $\pm$ 26.7 & 100.0 $\pm$ 0.0 & 100.0 $\pm$ 0.0 & 100.0 $\pm$ 0.0 & 82.1 $\pm$ 13.4 & 69.1 $\pm$ 14.2 \\
Pseudo-energy & 0.0 $\pm$ 0.0 & 22.7 $\pm$ 28.0 & 75.3 $\pm$ 28.6 & 100.0 $\pm$ 0.0 & 100.0 $\pm$ 0.0 & 100.0 $\pm$ 0.0 & 79.6 $\pm$ 11.0 & 89.8 $\pm$ 5.5 \\
CAP & 13.3 $\pm$ 13.6 & 17.3 $\pm$ 11.5 & 16.0 $\pm$ 17.0 & 40.7 $\pm$ 33.1 & 72.7 $\pm$ 27.5 & 94.7 $\pm$ 7.5 & 48.3 $\pm$ 15.5 & 67.5 $\pm$ 9.2 \\
PRADA & 40.7 $\pm$ 29.0 & 34.7 $\pm$ 30.9 & 32.0 $\pm$ 34.6 & 22.0 $\pm$ 22.7 & 11.3 $\pm$ 6.6 & 10.0 $\pm$ 4.3 & 22.0 $\pm$ 19.8 & 40.7 $\pm$ 9.9 \\
\bottomrule
\end{tabular}
\caption{Per-method detection results on MeaeQ-AGNews.}
\label{tab:per_dataset_meaeq_agnews}
\end{table*}

\begin{table*}[t]
\centering
\footnotesize
\setlength{\tabcolsep}{2.2pt}
\begin{tabular}{lcccccccc}
\toprule
Method & Benign FPR & 5\% TPR & 10\% TPR & 25\% TPR & 50\% TPR & 100\% TPR & Avg. TPR & Balanced Acc. \\
\midrule
MMD & 0.0 $\pm$ 0.0 & 94.7 $\pm$ 7.5 & 100.0 $\pm$ 0.0 & 100.0 $\pm$ 0.0 & 100.0 $\pm$ 0.0 & 100.0 $\pm$ 0.0 & 98.9 $\pm$ 1.5 & 99.5 $\pm$ 0.8 \\
Mahalanobis & 0.0 $\pm$ 0.0 & 100.0 $\pm$ 0.0 & 100.0 $\pm$ 0.0 & 100.0 $\pm$ 0.0 & 100.0 $\pm$ 0.0 & 100.0 $\pm$ 0.0 & 100.0 $\pm$ 0.0 & 100.0 $\pm$ 0.0 \\
DATE & 2.0 $\pm$ 1.6 & 12.0 $\pm$ 14.2 & 24.7 $\pm$ 29.2 & 56.0 $\pm$ 31.3 & 90.7 $\pm$ 13.2 & 100.0 $\pm$ 0.0 & 56.7 $\pm$ 16.4 & 77.3 $\pm$ 7.6 \\
SEAT & 41.3 $\pm$ 29.5 & 24.7 $\pm$ 33.5 & 58.7 $\pm$ 31.1 & 100.0 $\pm$ 0.0 & 100.0 $\pm$ 0.0 & 100.0 $\pm$ 0.0 & 76.7 $\pm$ 12.5 & 67.7 $\pm$ 20.9 \\
Pseudo-energy & 11.3 $\pm$ 17.9 & 98.0 $\pm$ 3.5 & 100.0 $\pm$ 0.0 & 100.0 $\pm$ 0.0 & 100.0 $\pm$ 0.0 & 100.0 $\pm$ 0.0 & 99.6 $\pm$ 0.7 & 94.1 $\pm$ 8.8 \\
CAP & 13.3 $\pm$ 14.8 & 1.3 $\pm$ 1.9 & 8.0 $\pm$ 5.9 & 54.7 $\pm$ 5.7 & 90.7 $\pm$ 11.8 & 93.3 $\pm$ 9.4 & 49.6 $\pm$ 4.1 & 68.1 $\pm$ 7.2 \\
PRADA & 1.3 $\pm$ 1.9 & 8.7 $\pm$ 8.4 & 14.0 $\pm$ 14.2 & 32.7 $\pm$ 17.5 & 36.0 $\pm$ 8.2 & 0.0 $\pm$ 0.0 & 18.3 $\pm$ 9.2 & 58.5 $\pm$ 3.7 \\
\bottomrule
\end{tabular}
\caption{Per-method detection results on MeaeQ-SST-2.}
\label{tab:per_dataset_meaeq_sst_2}
\end{table*}

\begin{table*}[t]
\centering
\footnotesize
\setlength{\tabcolsep}{2.2pt}
\begin{tabular}{lcccccccc}
\toprule
Method & Benign FPR & 5\% TPR & 10\% TPR & 25\% TPR & 50\% TPR & 100\% TPR & Avg. TPR & Balanced Acc. \\
\midrule
MMD & 1.3 $\pm$ 1.9 & 100.0 $\pm$ 0.0 & 100.0 $\pm$ 0.0 & 100.0 $\pm$ 0.0 & 100.0 $\pm$ 0.0 & 100.0 $\pm$ 0.0 & 100.0 $\pm$ 0.0 & 99.3 $\pm$ 0.9 \\
Mahalanobis & 12.7 $\pm$ 17.9 & 100.0 $\pm$ 0.0 & 100.0 $\pm$ 0.0 & 100.0 $\pm$ 0.0 & 100.0 $\pm$ 0.0 & 100.0 $\pm$ 0.0 & 100.0 $\pm$ 0.0 & 93.7 $\pm$ 9.0 \\
DATE & 11.3 $\pm$ 11.8 & 39.3 $\pm$ 43.5 & 37.3 $\pm$ 44.3 & 62.0 $\pm$ 42.9 & 87.3 $\pm$ 17.9 & 100.0 $\pm$ 0.0 & 65.2 $\pm$ 25.9 & 76.9 $\pm$ 18.0 \\
SEAT & 23.3 $\pm$ 27.5 & 37.3 $\pm$ 36.4 & 55.3 $\pm$ 40.5 & 100.0 $\pm$ 0.0 & 100.0 $\pm$ 0.0 & 100.0 $\pm$ 0.0 & 78.5 $\pm$ 14.9 & 77.6 $\pm$ 20.0 \\
Pseudo-energy & 33.3 $\pm$ 30.0 & 100.0 $\pm$ 0.0 & 100.0 $\pm$ 0.0 & 100.0 $\pm$ 0.0 & 100.0 $\pm$ 0.0 & 100.0 $\pm$ 0.0 & 100.0 $\pm$ 0.0 & 83.3 $\pm$ 15.0 \\
CAP & 6.0 $\pm$ 5.9 & 100.0 $\pm$ 0.0 & 100.0 $\pm$ 0.0 & 100.0 $\pm$ 0.0 & 100.0 $\pm$ 0.0 & 100.0 $\pm$ 0.0 & 100.0 $\pm$ 0.0 & 97.0 $\pm$ 2.9 \\
PRADA & 4.0 $\pm$ 4.3 & 100.0 $\pm$ 0.0 & 100.0 $\pm$ 0.0 & 100.0 $\pm$ 0.0 & 100.0 $\pm$ 0.0 & 100.0 $\pm$ 0.0 & 100.0 $\pm$ 0.0 & 98.0 $\pm$ 2.2 \\
\bottomrule
\end{tabular}
\caption{Per-method detection results on MeaeQ-IMDB.}
\label{tab:per_dataset_meaeq_imdb}
\end{table*}

\begin{table*}[t]
\centering
\footnotesize
\setlength{\tabcolsep}{2.2pt}
\begin{tabular}{lcccccccc}
\toprule
Method & Benign FPR & 5\% TPR & 10\% TPR & 25\% TPR & 50\% TPR & 100\% TPR & Avg. TPR & Balanced Acc. \\
\midrule
MMD & 0.7 $\pm$ 0.9 & 88.0 $\pm$ 4.3 & 100.0 $\pm$ 0.0 & 100.0 $\pm$ 0.0 & 100.0 $\pm$ 0.0 & 100.0 $\pm$ 0.0 & 97.6 $\pm$ 0.9 & 98.5 $\pm$ 0.1 \\
Mahalanobis & 3.3 $\pm$ 1.9 & 100.0 $\pm$ 0.0 & 100.0 $\pm$ 0.0 & 100.0 $\pm$ 0.0 & 100.0 $\pm$ 0.0 & 100.0 $\pm$ 0.0 & 100.0 $\pm$ 0.0 & 98.3 $\pm$ 0.9 \\
DATE & 6.7 $\pm$ 0.9 & 10.0 $\pm$ 11.3 & 25.3 $\pm$ 7.5 & 96.0 $\pm$ 0.0 & 100.0 $\pm$ 0.0 & 100.0 $\pm$ 0.0 & 66.3 $\pm$ 3.8 & 79.8 $\pm$ 1.4 \\
SEAT & 12.0 $\pm$ 3.3 & 51.3 $\pm$ 15.4 & 92.0 $\pm$ 5.7 & 100.0 $\pm$ 0.0 & 100.0 $\pm$ 0.0 & 100.0 $\pm$ 0.0 & 88.7 $\pm$ 4.2 & 88.3 $\pm$ 3.5 \\
Pseudo-energy & 7.3 $\pm$ 9.5 & 78.7 $\pm$ 10.3 & 100.0 $\pm$ 0.0 & 100.0 $\pm$ 0.0 & 100.0 $\pm$ 0.0 & 100.0 $\pm$ 0.0 & 95.7 $\pm$ 2.1 & 94.2 $\pm$ 3.9 \\
CAP & 14.7 $\pm$ 6.2 & 20.0 $\pm$ 17.0 & 42.7 $\pm$ 33.6 & 57.3 $\pm$ 31.7 & 85.3 $\pm$ 15.4 & 70.0 $\pm$ 42.4 & 55.1 $\pm$ 23.7 & 70.2 $\pm$ 10.1 \\
PRADA & 7.3 $\pm$ 4.1 & 10.7 $\pm$ 2.5 & 6.7 $\pm$ 0.9 & 16.0 $\pm$ 2.8 & 37.3 $\pm$ 5.2 & 100.0 $\pm$ 0.0 & 34.1 $\pm$ 0.8 & 63.4 $\pm$ 1.8 \\
\bottomrule
\end{tabular}
\caption{Per-method detection results on BERT-API-SST-2-Random.}
\label{tab:per_dataset_me_bert_sst2_random}
\end{table*}

\begin{table*}[t]
\centering
\footnotesize
\setlength{\tabcolsep}{2.2pt}
\begin{tabular}{lcccccccc}
\toprule
Method & Benign FPR & 5\% TPR & 10\% TPR & 25\% TPR & 50\% TPR & 100\% TPR & Avg. TPR & Balanced Acc. \\
\midrule
MMD & 0.7 $\pm$ 0.9 & 52.7 $\pm$ 5.7 & 100.0 $\pm$ 0.0 & 100.0 $\pm$ 0.0 & 100.0 $\pm$ 0.0 & 100.0 $\pm$ 0.0 & 90.5 $\pm$ 1.1 & 94.9 $\pm$ 0.8 \\
Mahalanobis & 3.3 $\pm$ 1.9 & 100.0 $\pm$ 0.0 & 100.0 $\pm$ 0.0 & 100.0 $\pm$ 0.0 & 100.0 $\pm$ 0.0 & 100.0 $\pm$ 0.0 & 100.0 $\pm$ 0.0 & 98.3 $\pm$ 0.9 \\
DATE & 6.7 $\pm$ 0.9 & 12.7 $\pm$ 6.6 & 18.0 $\pm$ 5.7 & 84.7 $\pm$ 3.8 & 100.0 $\pm$ 0.0 & 100.0 $\pm$ 0.0 & 63.1 $\pm$ 3.2 & 78.2 $\pm$ 1.1 \\
SEAT & 12.0 $\pm$ 3.3 & 52.0 $\pm$ 15.7 & 94.0 $\pm$ 5.7 & 100.0 $\pm$ 0.0 & 100.0 $\pm$ 0.0 & 100.0 $\pm$ 0.0 & 89.2 $\pm$ 4.3 & 88.6 $\pm$ 3.6 \\
Pseudo-energy & 7.3 $\pm$ 9.5 & 100.0 $\pm$ 0.0 & 100.0 $\pm$ 0.0 & 100.0 $\pm$ 0.0 & 100.0 $\pm$ 0.0 & 100.0 $\pm$ 0.0 & 100.0 $\pm$ 0.0 & 96.3 $\pm$ 4.7 \\
CAP & 14.7 $\pm$ 6.2 & 44.0 $\pm$ 16.1 & 76.0 $\pm$ 21.4 & 94.0 $\pm$ 8.5 & 98.0 $\pm$ 2.8 & 86.0 $\pm$ 19.8 & 79.6 $\pm$ 12.0 & 82.5 $\pm$ 7.4 \\
PRADA & 7.3 $\pm$ 4.1 & 27.3 $\pm$ 2.5 & 48.0 $\pm$ 8.6 & 68.7 $\pm$ 10.6 & 0.0 $\pm$ 0.0 & 100.0 $\pm$ 0.0 & 48.8 $\pm$ 3.1 & 70.7 $\pm$ 1.1 \\
\bottomrule
\end{tabular}
\caption{Per-method detection results on BERT-API-SST-2-Wiki.}
\label{tab:per_dataset_me_bert_sst2_wiki}
\end{table*}

\begin{table*}[t]
\centering
\footnotesize
\setlength{\tabcolsep}{2.2pt}
\begin{tabular}{lcccccccc}
\toprule
Method & Benign FPR & 5\% TPR & 10\% TPR & 25\% TPR & 50\% TPR & 100\% TPR & Avg. TPR & Balanced Acc. \\
\midrule
MMD & 0.7 $\pm$ 0.9 & 65.3 $\pm$ 7.5 & 100.0 $\pm$ 0.0 & 100.0 $\pm$ 0.0 & 100.0 $\pm$ 0.0 & 100.0 $\pm$ 0.0 & 93.1 $\pm$ 1.5 & 96.2 $\pm$ 1.1 \\
Mahalanobis & 4.7 $\pm$ 3.8 & 50.7 $\pm$ 10.6 & 96.0 $\pm$ 1.6 & 100.0 $\pm$ 0.0 & 100.0 $\pm$ 0.0 & 100.0 $\pm$ 0.0 & 89.3 $\pm$ 2.3 & 92.3 $\pm$ 0.9 \\
DATE & 6.0 $\pm$ 1.6 & 42.7 $\pm$ 22.3 & 76.0 $\pm$ 18.2 & 100.0 $\pm$ 0.0 & 100.0 $\pm$ 0.0 & 100.0 $\pm$ 0.0 & 83.7 $\pm$ 8.0 & 88.9 $\pm$ 4.3 \\
SEAT & 4.7 $\pm$ 3.4 & 8.7 $\pm$ 1.9 & 12.0 $\pm$ 1.6 & 92.7 $\pm$ 2.5 & 100.0 $\pm$ 0.0 & 100.0 $\pm$ 0.0 & 62.7 $\pm$ 0.5 & 79.0 $\pm$ 1.9 \\
Pseudo-energy & 7.3 $\pm$ 3.1 & 4.7 $\pm$ 1.2 & 1.3 $\pm$ 2.3 & 0.0 $\pm$ 0.0 & 0.0 $\pm$ 0.0 & 0.0 $\pm$ 0.0 & 1.2 $\pm$ 0.7 & 46.9 $\pm$ 1.3 \\
CAP & 11.3 $\pm$ 2.5 & 10.7 $\pm$ 5.2 & 21.3 $\pm$ 14.8 & 58.7 $\pm$ 38.8 & 76.0 $\pm$ 33.9 & 100.0 $\pm$ 0.0 & 53.3 $\pm$ 18.1 & 71.0 $\pm$ 8.0 \\
PRADA & 6.7 $\pm$ 4.1 & 6.7 $\pm$ 5.2 & 4.7 $\pm$ 2.5 & 3.3 $\pm$ 3.4 & 76.7 $\pm$ 3.4 & 21.3 $\pm$ 4.7 & 22.5 $\pm$ 2.5 & 57.9 $\pm$ 1.9 \\
\bottomrule
\end{tabular}
\caption{Per-method detection results on BERT-API-MNLI-Random.}
\label{tab:per_dataset_me_bert_mnli_random}
\end{table*}

\begin{table*}[t]
\centering
\footnotesize
\setlength{\tabcolsep}{2.2pt}
\begin{tabular}{lcccccccc}
\toprule
Method & Benign FPR & 5\% TPR & 10\% TPR & 25\% TPR & 50\% TPR & 100\% TPR & Avg. TPR & Balanced Acc. \\
\midrule
MMD & 0.7 $\pm$ 0.9 & 11.3 $\pm$ 3.8 & 100.0 $\pm$ 0.0 & 100.0 $\pm$ 0.0 & 100.0 $\pm$ 0.0 & 100.0 $\pm$ 0.0 & 82.3 $\pm$ 0.8 & 90.8 $\pm$ 0.8 \\
Mahalanobis & 4.7 $\pm$ 3.8 & 9.3 $\pm$ 5.7 & 21.3 $\pm$ 14.3 & 57.3 $\pm$ 12.3 & 98.7 $\pm$ 1.9 & 100.0 $\pm$ 0.0 & 57.3 $\pm$ 5.8 & 76.3 $\pm$ 1.4 \\
DATE & 6.7 $\pm$ 1.9 & 3.3 $\pm$ 0.9 & 8.7 $\pm$ 4.7 & 40.7 $\pm$ 23.5 & 66.7 $\pm$ 34.7 & 90.7 $\pm$ 13.2 & 42.0 $\pm$ 15.4 & 67.7 $\pm$ 6.7 \\
SEAT & 4.7 $\pm$ 3.4 & 15.3 $\pm$ 2.5 & 28.0 $\pm$ 4.3 & 99.3 $\pm$ 0.9 & 100.0 $\pm$ 0.0 & 100.0 $\pm$ 0.0 & 68.5 $\pm$ 1.5 & 81.9 $\pm$ 2.0 \\
Pseudo-energy & 7.3 $\pm$ 3.1 & 12.0 $\pm$ 6.9 & 22.7 $\pm$ 11.0 & 73.3 $\pm$ 10.3 & 99.3 $\pm$ 1.2 & 100.0 $\pm$ 0.0 & 61.5 $\pm$ 5.4 & 77.1 $\pm$ 1.4 \\
CAP & 11.3 $\pm$ 2.5 & 12.0 $\pm$ 6.5 & 18.0 $\pm$ 9.9 & 45.3 $\pm$ 24.1 & 83.3 $\pm$ 23.6 & 100.0 $\pm$ 0.0 & 51.7 $\pm$ 11.7 & 70.2 $\pm$ 4.8 \\
PRADA & 6.7 $\pm$ 4.1 & 2.7 $\pm$ 2.5 & 2.7 $\pm$ 2.5 & 2.0 $\pm$ 1.6 & 11.3 $\pm$ 0.9 & 63.3 $\pm$ 10.6 & 16.4 $\pm$ 2.0 & 54.9 $\pm$ 1.2 \\
\bottomrule
\end{tabular}
\caption{Per-method detection results on BERT-API-MNLI-Wiki.}
\label{tab:per_dataset_me_bert_mnli_wiki}
\end{table*}

\begin{table*}[t]
\centering
\footnotesize
\setlength{\tabcolsep}{2.2pt}
\begin{tabular}{lcccccccc}
\toprule
Method & Benign FPR & 5\% TPR & 10\% TPR & 25\% TPR & 50\% TPR & 100\% TPR & Avg. TPR & Balanced Acc. \\
\midrule
MMD & 0.0 $\pm$ 0.0 & 86.0 $\pm$ 4.3 & 100.0 $\pm$ 0.0 & 100.0 $\pm$ 0.0 & 100.0 $\pm$ 0.0 & 100.0 $\pm$ 0.0 & 97.2 $\pm$ 0.9 & 98.6 $\pm$ 0.4 \\
Mahalanobis & 15.3 $\pm$ 3.8 & 89.3 $\pm$ 8.1 & 100.0 $\pm$ 0.0 & 100.0 $\pm$ 0.0 & 100.0 $\pm$ 0.0 & 100.0 $\pm$ 0.0 & 97.9 $\pm$ 1.6 & 91.3 $\pm$ 1.1 \\
DATE & 8.7 $\pm$ 2.5 & 100.0 $\pm$ 0.0 & 100.0 $\pm$ 0.0 & 100.0 $\pm$ 0.0 & 100.0 $\pm$ 0.0 & 100.0 $\pm$ 0.0 & 100.0 $\pm$ 0.0 & 95.7 $\pm$ 1.2 \\
SEAT & 18.0 $\pm$ 5.9 & 13.3 $\pm$ 6.2 & 88.0 $\pm$ 6.5 & 100.0 $\pm$ 0.0 & 100.0 $\pm$ 0.0 & 100.0 $\pm$ 0.0 & 80.3 $\pm$ 0.5 & 81.1 $\pm$ 2.8 \\
Pseudo-energy & 26.0 $\pm$ 17.3 & 35.3 $\pm$ 18.9 & 50.7 $\pm$ 20.2 & 72.0 $\pm$ 21.6 & 99.3 $\pm$ 1.2 & 100.0 $\pm$ 0.0 & 71.5 $\pm$ 11.6 & 72.7 $\pm$ 3.8 \\
CAP & 14.0 $\pm$ 2.8 & 10.7 $\pm$ 0.9 & 17.3 $\pm$ 4.7 & 62.7 $\pm$ 9.6 & 99.3 $\pm$ 0.9 & 100.0 $\pm$ 0.0 & 58.0 $\pm$ 3.1 & 72.0 $\pm$ 2.6 \\
PRADA & 5.3 $\pm$ 4.1 & 9.3 $\pm$ 3.8 & 19.3 $\pm$ 3.4 & 78.0 $\pm$ 4.9 & 99.3 $\pm$ 0.9 & 100.0 $\pm$ 0.0 & 61.2 $\pm$ 0.3 & 77.9 $\pm$ 2.0 \\
\bottomrule
\end{tabular}
\caption{Per-method detection results on BERT-API-SQuAD-Random.}
\label{tab:per_dataset_me_bert_squad_random}
\end{table*}

\begin{table*}[t]
\centering
\footnotesize
\setlength{\tabcolsep}{2.2pt}
\begin{tabular}{lcccccccc}
\toprule
Method & Benign FPR & 5\% TPR & 10\% TPR & 25\% TPR & 50\% TPR & 100\% TPR & Avg. TPR & Balanced Acc. \\
\midrule
MMD & 0.0 $\pm$ 0.0 & 4.0 $\pm$ 4.3 & 60.7 $\pm$ 5.7 & 100.0 $\pm$ 0.0 & 100.0 $\pm$ 0.0 & 100.0 $\pm$ 0.0 & 72.9 $\pm$ 1.9 & 86.5 $\pm$ 1.0 \\
Mahalanobis & 15.3 $\pm$ 3.8 & 26.0 $\pm$ 13.4 & 36.0 $\pm$ 15.6 & 61.3 $\pm$ 19.1 & 88.0 $\pm$ 9.1 & 100.0 $\pm$ 0.0 & 62.3 $\pm$ 10.9 & 73.5 $\pm$ 3.6 \\
DATE & 8.7 $\pm$ 2.5 & 30.7 $\pm$ 13.2 & 68.7 $\pm$ 13.2 & 100.0 $\pm$ 0.0 & 100.0 $\pm$ 0.0 & 100.0 $\pm$ 0.0 & 79.9 $\pm$ 5.3 & 85.6 $\pm$ 1.9 \\
SEAT & 18.0 $\pm$ 5.9 & 18.0 $\pm$ 9.9 & 19.3 $\pm$ 10.4 & 12.7 $\pm$ 5.0 & 70.7 $\pm$ 9.3 & 100.0 $\pm$ 0.0 & 44.1 $\pm$ 3.0 & 63.1 $\pm$ 1.6 \\
Pseudo-energy & 26.0 $\pm$ 17.3 & 20.0 $\pm$ 16.4 & 17.3 $\pm$ 16.7 & 2.7 $\pm$ 2.3 & 0.0 $\pm$ 0.0 & 0.0 $\pm$ 0.0 & 8.0 $\pm$ 6.8 & 41.0 $\pm$ 6.0 \\
CAP & 14.0 $\pm$ 2.8 & 12.7 $\pm$ 1.9 & 6.7 $\pm$ 3.4 & 16.7 $\pm$ 9.6 & 32.0 $\pm$ 8.2 & 98.7 $\pm$ 1.9 & 33.3 $\pm$ 2.9 & 59.7 $\pm$ 1.4 \\
PRADA & 5.3 $\pm$ 4.1 & 11.3 $\pm$ 0.9 & 22.0 $\pm$ 1.6 & 82.0 $\pm$ 5.7 & 100.0 $\pm$ 0.0 & 100.0 $\pm$ 0.0 & 63.1 $\pm$ 1.2 & 78.9 $\pm$ 2.2 \\
\bottomrule
\end{tabular}
\caption{Per-method detection results on BERT-API-SQuAD-Wiki.}
\label{tab:per_dataset_me_bert_squad_wiki}
\end{table*}

\begin{table*}[t]
\centering
\footnotesize
\setlength{\tabcolsep}{2.2pt}
\begin{tabular}{lcccccccc}
\toprule
Method & Benign FPR & 5\% TPR & 10\% TPR & 25\% TPR & 50\% TPR & 100\% TPR & Avg. TPR & Balanced Acc. \\
\midrule
MMD & 0.0 $\pm$ 0.0 & 90.7 $\pm$ 10.5 & 100.0 $\pm$ 0.0 & 100.0 $\pm$ 0.0 & 100.0 $\pm$ 0.0 & 100.0 $\pm$ 0.0 & 98.1 $\pm$ 2.1 & 99.1 $\pm$ 1.0 \\
Mahalanobis & 34.7 $\pm$ 46.2 & 92.0 $\pm$ 9.9 & 100.0 $\pm$ 0.0 & 100.0 $\pm$ 0.0 & 100.0 $\pm$ 0.0 & 100.0 $\pm$ 0.0 & 98.4 $\pm$ 2.0 & 81.9 $\pm$ 22.5 \\
DATE & 17.3 $\pm$ 12.0 & 100.0 $\pm$ 0.0 & 100.0 $\pm$ 0.0 & 100.0 $\pm$ 0.0 & 100.0 $\pm$ 0.0 & 100.0 $\pm$ 0.0 & 100.0 $\pm$ 0.0 & 91.3 $\pm$ 6.0 \\
SEAT & 0.0 $\pm$ 0.0 & 8.7 $\pm$ 4.1 & 100.0 $\pm$ 0.0 & 100.0 $\pm$ 0.0 & 100.0 $\pm$ 0.0 & 100.0 $\pm$ 0.0 & 81.7 $\pm$ 0.8 & 90.9 $\pm$ 0.4 \\
Pseudo-energy & 30.7 $\pm$ 53.1 & 83.3 $\pm$ 16.0 & 100.0 $\pm$ 0.0 & 100.0 $\pm$ 0.0 & 100.0 $\pm$ 0.0 & 100.0 $\pm$ 0.0 & 96.7 $\pm$ 3.2 & 83.0 $\pm$ 25.1 \\
CAP & 26.0 $\pm$ 34.0 & 31.3 $\pm$ 44.3 & 54.7 $\pm$ 40.3 & 86.0 $\pm$ 19.8 & 100.0 $\pm$ 0.0 & 100.0 $\pm$ 0.0 & 74.4 $\pm$ 19.2 & 74.2 $\pm$ 9.0 \\
PRADA & 3.3 $\pm$ 1.9 & 10.0 $\pm$ 12.8 & 26.0 $\pm$ 27.0 & 91.3 $\pm$ 6.2 & 100.0 $\pm$ 0.0 & 100.0 $\pm$ 0.0 & 65.5 $\pm$ 9.1 & 81.1 $\pm$ 3.6 \\
\bottomrule
\end{tabular}
\caption{Per-method detection results on BERT-API-BoolQ-Random.}
\label{tab:per_dataset_me_bert_boolq_random}
\end{table*}

\begin{table*}[t]
\centering
\footnotesize
\setlength{\tabcolsep}{2.2pt}
\begin{tabular}{lcccccccc}
\toprule
Method & Benign FPR & 5\% TPR & 10\% TPR & 25\% TPR & 50\% TPR & 100\% TPR & Avg. TPR & Balanced Acc. \\
\midrule
MMD & 0.0 $\pm$ 0.0 & 4.0 $\pm$ 4.3 & 51.3 $\pm$ 26.4 & 100.0 $\pm$ 0.0 & 100.0 $\pm$ 0.0 & 100.0 $\pm$ 0.0 & 71.1 $\pm$ 5.9 & 85.5 $\pm$ 3.0 \\
Mahalanobis & 34.7 $\pm$ 46.2 & 38.0 $\pm$ 44.2 & 41.3 $\pm$ 42.0 & 68.0 $\pm$ 25.4 & 98.7 $\pm$ 1.9 & 100.0 $\pm$ 0.0 & 69.2 $\pm$ 22.4 & 67.3 $\pm$ 12.5 \\
DATE & 14.7 $\pm$ 9.6 & 54.0 $\pm$ 33.3 & 100.0 $\pm$ 0.0 & 100.0 $\pm$ 0.0 & 100.0 $\pm$ 0.0 & 100.0 $\pm$ 0.0 & 90.8 $\pm$ 6.7 & 88.1 $\pm$ 2.1 \\
SEAT & 0.0 $\pm$ 0.0 & 7.3 $\pm$ 3.4 & 45.3 $\pm$ 5.7 & 100.0 $\pm$ 0.0 & 99.3 $\pm$ 0.9 & 100.0 $\pm$ 0.0 & 70.4 $\pm$ 1.5 & 85.2 $\pm$ 0.7 \\
Pseudo-energy & 30.7 $\pm$ 53.1 & 29.3 $\pm$ 50.8 & 28.7 $\pm$ 49.7 & 24.0 $\pm$ 41.6 & 5.3 $\pm$ 9.2 & 0.0 $\pm$ 0.0 & 17.5 $\pm$ 30.3 & 43.4 $\pm$ 11.4 \\
CAP & 26.0 $\pm$ 34.0 & 23.3 $\pm$ 27.4 & 23.3 $\pm$ 26.0 & 36.0 $\pm$ 19.3 & 74.7 $\pm$ 16.0 & 93.3 $\pm$ 9.4 & 50.1 $\pm$ 16.1 & 62.1 $\pm$ 9.0 \\
PRADA & 3.3 $\pm$ 1.9 & 12.0 $\pm$ 12.8 & 26.7 $\pm$ 30.7 & 92.7 $\pm$ 5.2 & 100.0 $\pm$ 0.0 & 100.0 $\pm$ 0.0 & 66.3 $\pm$ 9.7 & 81.5 $\pm$ 3.9 \\
\bottomrule
\end{tabular}
\caption{Per-method detection results on BERT-API-BoolQ-Wiki.}
\label{tab:per_dataset_me_bert_boolq_wiki}
\end{table*}

\clearpage

\end{document}